\documentclass[aps,prd,onecolumn,showpacs,floatfix,superscriptaddress]{revtex4-2}
\usepackage{graphicx}
\usepackage{dsfont}
\usepackage{xcolor}
\usepackage{multirow}
\usepackage[normalem]{ulem}
\usepackage{enumitem} 
\usepackage{lineno}
\usepackage{bm}
\definecolor{refcol}{RGB}{34,34,178}
\usepackage[colorlinks,linkcolor=blue,citecolor=blue,urlcolor=blue]{hyperref}
\usepackage{microtype}
\usepackage{float}
\usepackage[caption = false]{subfig}
\usepackage{appendix}
\usepackage{multirow}
\usepackage{soul}
\usepackage{appendix}
\usepackage{lineno}
\graphicspath{{./Fig/}}

\definecolor{DB}{RGB}{178,34,34}
 
\begin{document}
\title{Chiral condensate and the equation of state at nonzero baryon density
    from the hadron resonance gas model with a repulsive mean field}
\author{Deeptak Biswas}
\affiliation{The Institute of Mathematical Sciences, a CI of 
Homi Bhabha National Institute, Chennai, 600113, India}
\affiliation{School of Physical Sciences, National Institute of Science 
Education and Research, An OCC of Homi Bhabha National Institute, Jatni-752050, 
India} 
\author{Peter Petreczky}
\affiliation{Physics Department, Brookhaven National Laboratory, Upton 
NY 11973, USA}
\author{Sayantan Sharma}
\affiliation{The Institute of Mathematical Sciences, a CI of 
Homi Bhabha National Institute, Chennai, 600113, India}
%
\begin{abstract}
We study the QCD equation of state and the chiral condensate using 
the hadron resonance gas model with repulsive mean-field interactions. We find that
the repulsive interactions improve the agreement with 
the lattice results on the derivatives of the pressure with respect
to the baryon chemical potential up to eighth order. 
From the temperature dependence of the chiral
condensate we estimate the crossover temperature as a function 
of baryon chemical potential, $T_{pc}(\mu_B)$. We find that the chiral crossover line starts
to deviate significantly from the chemical freeze-out line already for $\mu_B>400$ MeV.
Furthermore, we find that the chiral pseudocritical line can be parametrized as 
$T_{pc}(\mu_B)/T_{pc}(0)=1-\kappa_2 (\mu_B/T_{pc} (0))^2-\kappa_4 (\mu_B/T_{pc} (0))^4$
with $\kappa_2=0.0150(2)$ and $\kappa_4=3.1(6) \times 10^{-5}$, which are in agreement
with lattice QCD results for small values of $\mu_B$. For the first time 
we find a tiny but non-zero value of $\kappa_4$ in our study.
\end{abstract}

%
\maketitle
%

\section{Introduction}
Understanding QCD matter at nonzero temperatures
and baryon densities presents several challenges 
on the theory side. Apart from asymptotically large temperatures 
and baryon densities when perturbation theory can make predictions 
for thermodynamics, the problem is nonperturbative.  For zero baryon 
density, lattice QCD methods have provided continuum extrapolated
results for physical quark masses for many thermodynamic quantities.
However, due to the infamous \emph{sign~problem}, the lattice 
Monte Carlo techniques based on the method of important sampling 
break down at finite densities. Attempts to extend the lattice QCD 
calculations to nonzero baryon densities through 
Taylor expansion in baryon chemical potential, $\mu_B$, 
or the imaginary chemical potential approach has allowed us to 
understand QCD thermodynamics $\mu_B<2.5~T$~\cite{Bazavov:2017dus, 
Borsanyi:2018grb, DElia:2016jqh, Bollweg:2022rps}, 
which can be extended in a certain temperature regime to $\mu_B/T \approx 
3.5$~\cite{Borsanyi:2021sxv}. 

Below the chiral crossover temperature, QCD thermodynamics can be fairly well understood
using the hadron resonance gas (HRG) model. The main idea of this model is that 
the interactions between hadrons can be taken into account through hadronic
resonances and the interacting gas of hadrons can be replaced by a gas of 
noninteracting hadrons and hadronic resonances treated as stable particles. 
This approach can be justified within the framework of relativistic virial 
expansion~\cite{Dashen:1969ep,Venugopalan:1992hy} and the comparison
between the HRG model results and the lattice data~\cite{Karsch:2003vd, Karsch:2003zq, Ejiri:2005wq,Huovinen:2009yb,Borsanyi:2010bp,HotQCD:2012fhj}.
Indeed, the most recent lattice QCD calculations of the QCD pressure and
the second order fluctuations and correlations of conserved charges agree 
well with the  the HRG model results ~\cite{Borsanyi:2013bia,Bazavov:2014pvz,Bazavov:2017dsy,Bellwied:2015lba,Borsanyi:2018grb,Bollweg:2021vqf,Borsanyi:2013bia,Bazavov:2014pvz,Bazavov:2017dsy}. 
For some quantities like baryon-strangeness or baryon-charm correlations, it is important
to include additional hadron states that are not listed by the Particle Data Group (PDG)
but are predicted by lattice QCD and quark models ~\cite{Bazavov:2014xya,Bazavov:2014yba,Bellwied:2019pxh,Bollweg:2021vqf}.
For strangeness-baryon number correlation, this idea is further supported by 
the calculation within the relativistic virial expansion with state-of-the-art 
phase shift analysis~\cite{Fernandez-Ramirez:2018vzu}.

The temperature dependence of the chiral condensate has been
studied within the HRG model ~\cite{Gerber:1988tt,Toublan:2004ks,Jankowski:2012ms,Andersen:2022clu,Biswas:2022vat},
and attempts to use the HRG model to estimate the chiral crossover temperature at small 
baryon density have been made recently~\cite{Biswas:2022vat}. There is a widespread 
expectation that the chiral crossover may become a true phase transition at large baryon density,
namely that the crossover will turn into a first-order phase transition at $\mu_B=\mu_B^{CEP}$
corresponding to the critical end-point (CEP). Lattice QCD results strongly disfavor 
$\mu_B^{CEP}<400$ MeV for a range of temperatures around $0.85~T_{pc}$, while functional 
renormalization group studies estimate $\mu_B^{CEP}$ to be even larger, around 
$635$ MeV \cite{Fu:2019hdw} at $T=107$ MeV.
Therefore, it would be interesting to estimate the chiral crossover temperature
at larger values of baryon density using the HRG model framework.

The agreement between lattice QCD and the HRG model results no longer holds
for higher-order fluctuations of conserved charges ~\cite{DElia:2016jqh,Borsanyi:2018grb,Bollweg:2022rps,Bollweg:2022fqq}.
This may highlight the importance of the inclusion of non-resonant attractive 
interactions as well as repulsive interactions among the baryons, which could 
become very important at higher baryon densities. The most common approach to 
include repulsive interactions within the HRG model is through the inclusion 
of a hard-core repulsive excluded volume for hadrons~\cite{Rischke:1991ke, Cleymans:1992jz, Yen:1997rv, Tiwari:2011km, Bhattacharyya:2013oya, Albright:2014gva, Kadam:2015xsa, Kadam:2015dda}. 
Recently these models were refined to include both repulsive and attractive
nonresonant interactions via a van der Waals-like potential~
\cite{Vovchenko:2016rkn, Vovchenko:2017zpj, Samanta:2017yhh, Sarkar:2018mbk, Vovchenko:2020lju}.
Another approach to include the repulsive interactions is through the 
mean-field approximation. Here the strength of the repulsive interaction 
is proportional to the baryon density~\cite{Olive:1980dy, Olive:1982we, Kapusta:1982qd}. 
It has been shown that this model can explain the deviations from
the HRG model seen in the lattice QCD calculations of higher-order
derivatives of the QCD pressure with respect to the baryon chemical 
potential~\cite{Huovinen:2017ogf}. Following this work the use of the HRG model 
with repulsive mean-field interactions was explored to explain 
the lattice results on the QCD equation of state~\cite{Huovinen:2018xmq, Pal:2021qav} 
and fluctuations and correlations of conserved charges~\cite{Pal:2020ucy, Pal:2023zzp}.

The aim of this paper is to study the temperature and the $\mu_B$ dependence of the chiral
condensate, and using this to estimate the chiral crossover temperature as a function
of $\mu_B$. We will also revisit the description of the higher-order baryon number 
fluctuations and the QCD equation of state within the HRG model with repulsive mean-field 
interactions in light of the newest lattice QCD results.
The paper is organized as follows: In the next section, we review the HRG model
with repulsive interactions. In that section we also present the comparison
of this model with the state-of-the-art lattice QCD results on fluctuations of
conserved charges. In Sec.~\ref{secIII} we discuss the calculation of the chiral condensate
as a function of temperature and baryon chemical potential. In Sec.~\ref{secIV} we present our 
results, i.e., the estimate of the chiral crossover temperature as a function of baryon 
chemical potential, and the curvature of the pseudocritical line. In Sec.~\ref{secV}, we discuss the 
temperature and baryon density dependence of the speed of sound for the strangeness neutral 
system. Finally, our conclusions are presented in Sec.~\ref{secVI}. In the Appendices, we give some 
technical details of the calculations.

\section{Including Repulsive baryon interactions within the HRG model}\label{secII}

It is well known from nucleon-nucleon scattering experiments that there are repulsive 
interactions between nucleons at short distances (or equivalently at high energies).
From the lattice results we know that this is also true for other baryons \cite{Aoki:2023qih}.
These interactions will become important at large baryon density.
Since we are interested in extending the HRG model to high baryon density 
the effect of the repulsive baryon-baryon interactions has to be included.
The repulsive baryon-baryon interactions could play a role also at zero or small
baryon densities as the temperature increases toward $T_{pc}$, since the 
abundance of baryons and antibaryons increases.
But from
previous analysis we know that the corresponding effect is small for
the pressure and energy density \cite{Huovinen:2018xmq}. However, higher-order derivatives of the pressure with respect to the baryon
chemical potential are sensitive to the effect of the repulsive interactions
even at zero baryon density \cite{Huovinen:2017ogf}. This is because these
derivatives carry information about the physics at high baryon density even when evaluated
at zero baryon chemical potential. Comparison of HRG calculations with the lattice
results on higher-order derivatives of the pressure with respect to baryon chemical
potential allows one to constrain the contribution of the repulsive interactions.
As already mentioned in the introduction in this work
we use the repulsive mean-field to include the effect of repulsive baryon-baryon interactions
in the HRG model. 

The pressure for an interacting ensemble of baryons and anti-baryons 
with densities $n_b$ and $n_{\bar{b}}$ respectively can be represented
within the repulsive mean-field model as~\cite{Pal:2020ucy, Pal:2021qav}
\begin{equation}
P^{B\{\bar{B}\}}_{int}=T\sum \limits_{i\in B \{\bar B \}} \int 
g_i \frac{d^3 p}{(2 \pi)^3} \text{ln}\bigg[1+ e^{-\beta(E_i-\mu_{eff}})\bigg] 
+\frac{K}{2} n_{b\{\bar{b}\}}^2.
\label{Eq.Pint}
\end{equation}
Here the effective chemical potential for the $i$th species is defined as 
$\mu_{eff}=B_i \mu_B - K n_{b\{\bar{b}\}}$, with $B_i$ being the 
baryon number. $B_i=+1$ for baryons, and $B_i=-1$ for antibaryons respectively, 
and $\beta=1/T$, where $T$ is the temperature of this system.
The number densities for baryons and anti-baryons can be solved 
self-consistently from the following pair of equations:
\begin{equation}
n_{{b}}=\sum_{i\in B}\int g_i \frac{d^3 p}{(2 \pi)^3}\:\frac{1}{e^{\beta{(E_{i}-\mu_B+K n_b)}}+1}
~~~{\text{and}}~~~
n_{{\bar{b}}}=\sum_{i\in \bar{B}}\int g_i \frac{d^3 p}{(2 \pi)^3}\:\frac{1}{e^{\beta{(E_{i}+\mu_B+K n_{\bar{b}})}}+1}.
\label{Eq.nb}
\end{equation}
The total pressure within the HRG model is the sum of the contributions 
from this interacting ensemble of the (anti)baryons and the 
noninteracting ensemble of mesons. We include the quark model (QM) predicted 
states~\cite{Capstick:1986ter, Ebert:2009ub} in addition to the 
Particle Data Group (PDG) listed hadrons up to mass $3$ GeV 
following~\cite{Alba:2017mqu, Alba:2020jir}. 
The PDG list was made considering all the states, even the ones with large uncertainties, 
and to avoid double counting the QM states are replaced by the experimentally determined 
states if they have the same mass and quantum numbers.
We denote this HRG model with the extended list of particles
as the QMHRG model. For $\mu_B=0$ we can use the Boltzmann approximation for baryons 
and antibaryons due to their large masses. Therefore, for the pressure 
and the number density we could write the following set of equations~\cite{Huovinen:2017ogf}:
\begin{equation}
P_{int}=T (n_b + n_{\bar{b}}) + \frac{K}{2}(n_{b}^2 + n_{\bar{b}}^2) 
~~~~\text{and}~~~~ 
n_{b\{\bar{b}\}}=\sum_{i=B\{\bar{B}\}} \int g_i \frac{d^3 p}{(2 \pi)^3} 
e^{-\beta(E_i-\mu_{eff})} 
\end{equation}
This approximation greatly simplifies the calculations of the baryon number
susceptibilities, which are defined as
\begin{equation}
\chi_{B}^{n}=\frac{\partial^n \left[P(\mu_B/T)/T^4\right]}{\partial (\mu_B/T)^n },~~~n=2,4,6,8
\end{equation}
The explicit expressions of $\chi_n^B$ are presented in Appendix \ref{app:fluct}. 
In this work, we have included the partial pressures of the QMHRG states
with repulsive interactions among the (anti) baryons at the mean-field 
level. Typically in the earlier works the mean-field coefficient has been 
chosen to be $K=450$ MeV $\text{fm}^3$ which is equivalent to $56.25~\text{GeV}^{-2}$~
\cite{Sollfrank:1996hd,Huovinen:2017ogf} from phenomenological considerations. 
Here we suggest a new way of estimating this constant.
\begin{figure}
\includegraphics[width=8.8cm]{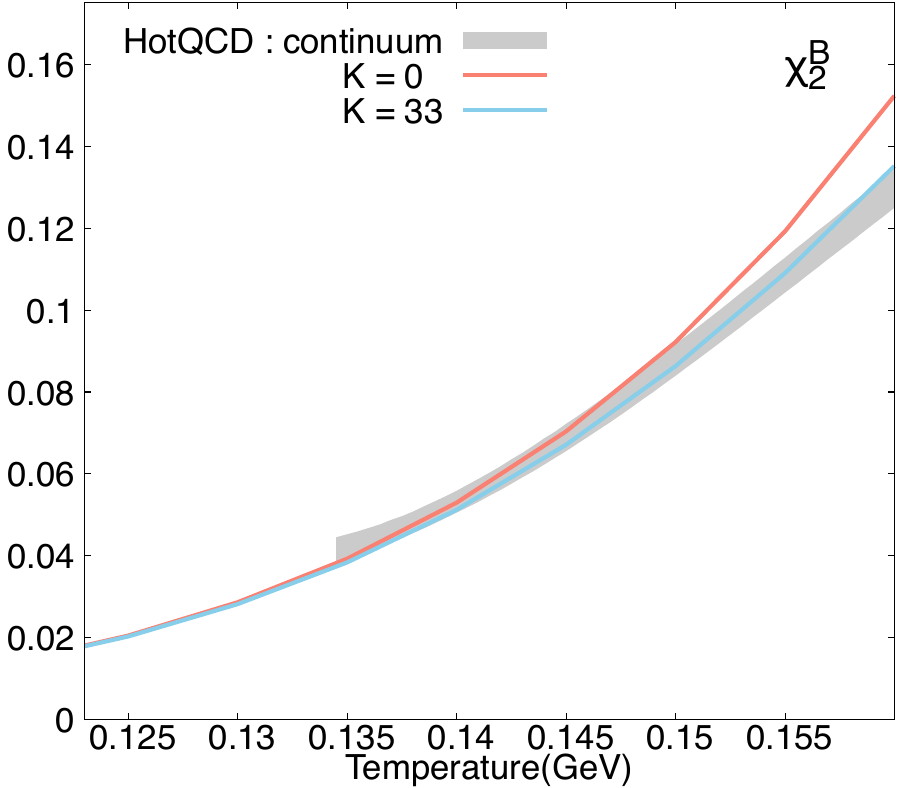}
\includegraphics[width=8.8cm]{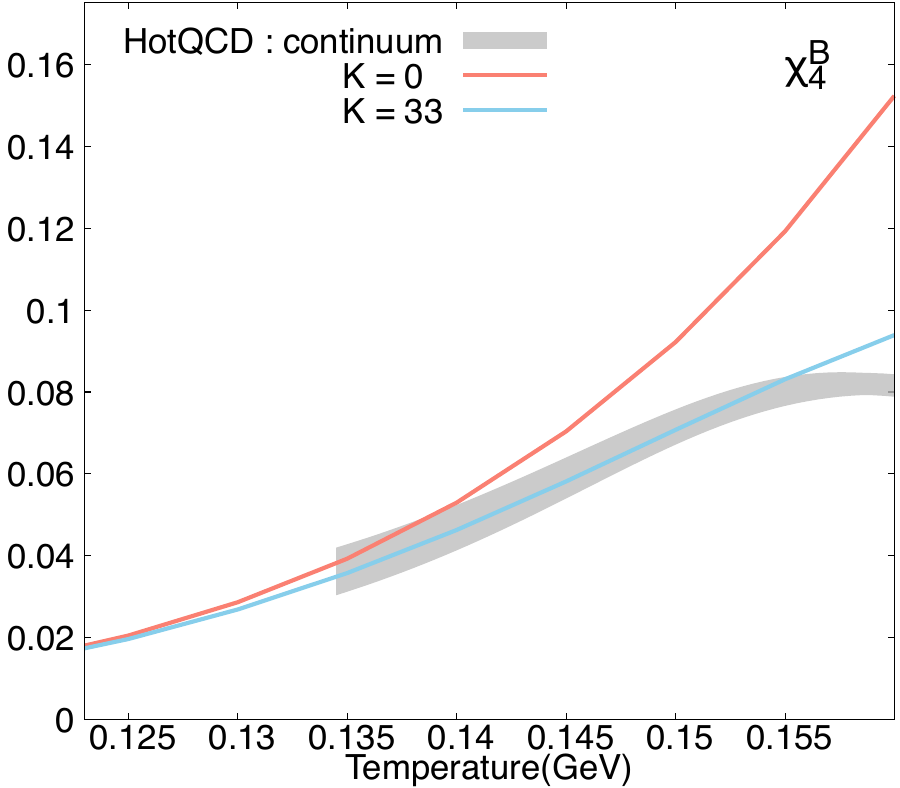}\\
\includegraphics[width=8.8cm]{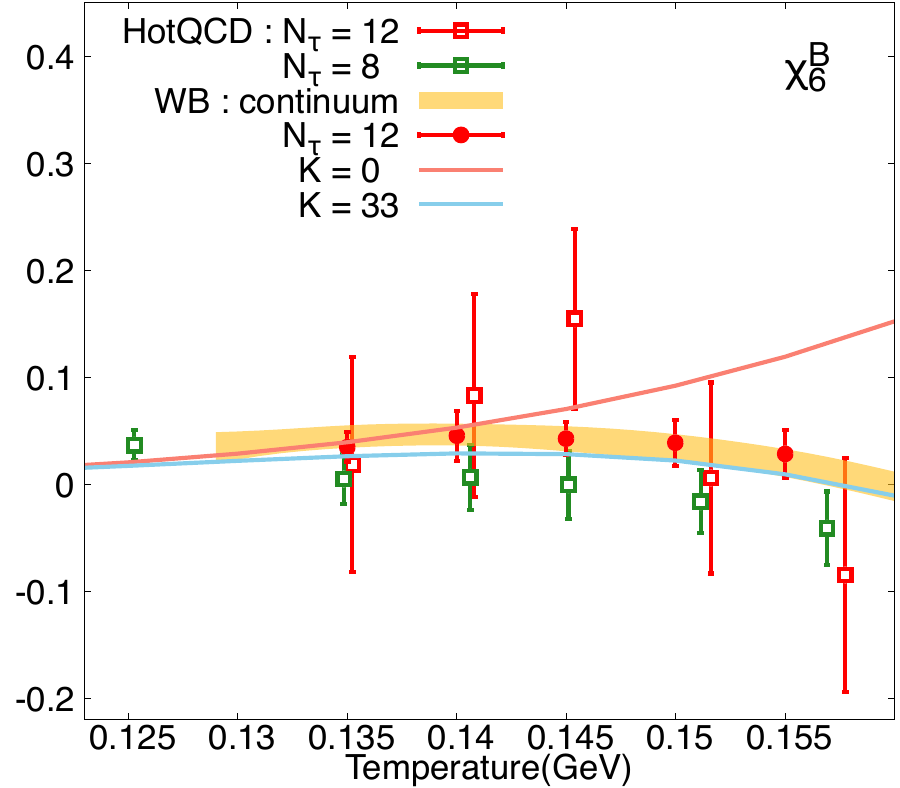}
\includegraphics[width=8.8cm]{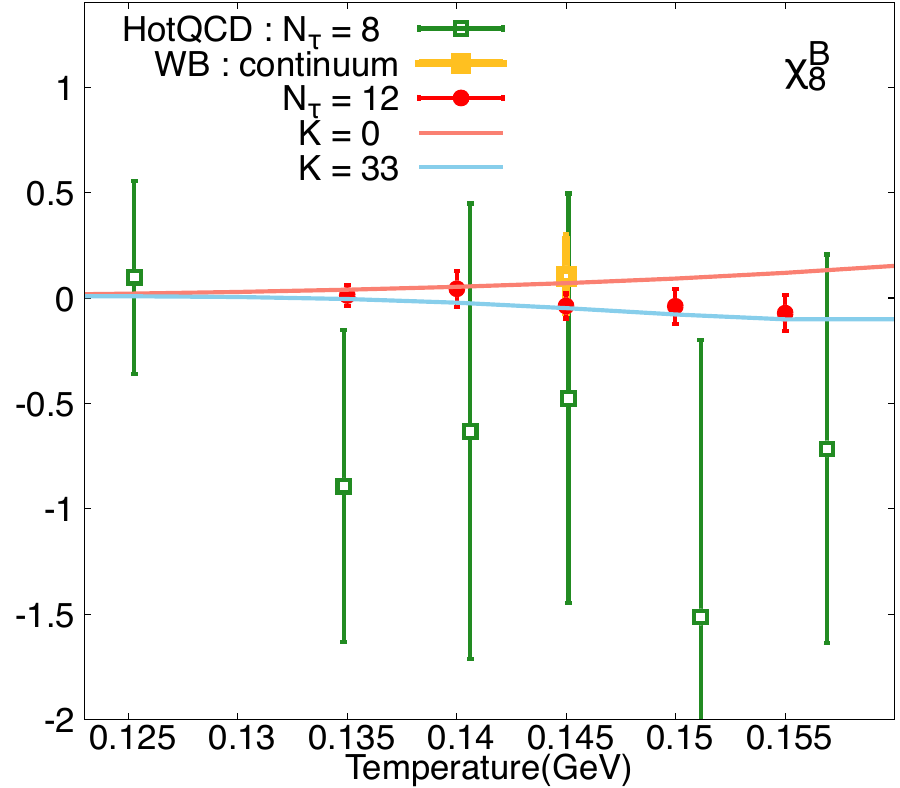}
\caption{The second (top left), fourth (top right), sixth (bottom left), 
and eighth (bottom right) order baryon number fluctuations results compared 
between ideal QMHRG (red line) and mean-field  QMHRG (blue line) models
and lattice QCD results with physical quark masses. 
Lattice results for $N_\tau = 12$ and $N_\tau = 8$ are 
shown as red and green points respectively with the squares representing 
HotQCD~\cite{Bazavov:2020bjn} results and the circles corresponding to 
Wuppertal-Budapest (WB) results~\cite{Borsanyi:2018grb}. The gray and yellow 
bands represent the fluctuation data in the continuum limit from  
HotQCD~\cite{Bazavov:2020bjn} and WB~\cite{Borsanyi:2023wno} Collaborations.}
\label{fig:chi2468B}
\end{figure}
Our approach is motivated by the fact that we have high
precision lattice QCD data for $\chi_2^B$ and $\chi_4^B$ 
extrapolated to the continuum limit~\cite{Bollweg:2021vqf,Bollweg:2022rps,Bollweg:2022fqq}.
Therefore, we adjust the value of $K$ to reproduce these lattice QCD results.
In Fig. \ref{fig:chi2468B} we show the comparison of our results for the second- and fourth-order
baryon number fluctuations with the corresponding lattice QCD results.
The lattice QCD results on $\chi_{2}^{B}$ results disagree with the QMHRG model results for 
$T>150$ MeV, while the lattice QCD results for $\chi_{4}^{B}$ start to disagree with 
QMHRG model results already for $T>140$ MeV, when no repulsive interactions are taken 
into account. By including repulsive mean-field interactions within the QMHRG model with 
$K=33~\text{GeV}^{-2}$, we obtain a good agreement with the lattice QCD results for
both $\chi_{2}^{B}$ and $\chi_4^B$ up to $T \approx 155$ MeV.
In addition there are lattice QCD results for $N_{\tau}=8$ and $N_{\tau}=12$ from the 
HotQCD Collaboration~\cite{Bazavov:2020bjn} and $N_{\tau}=12$ results from the 
Wuppertal-Budapest Collaboration~\cite{Borsanyi:2018grb} for spatial volume  $V^{1/3}T=4$.
Here $N_{\tau}$ is the temporal lattice extent that is related to the temperature and the
lattice spacing $a$, as $N_{\tau}=1/(a T)$. 
Very recently continuum extrapolated lattice QCD results on 
for sixth- and eighth-order baryon number fluctuations were released for small spatial volume $V^{1/3}T=2$ \cite{Borsanyi:2023wno}.

In the lower panels of Fig. \ref{fig:chi2468B} 
we show the comparison of the QMHRG model results with the corresponding lattice QCD results 
for $\chi_6^B$ and $\chi_8^B$. We see from the figure that the QMHRG model results with repulsive 
mean-field and $K=33~\text{GeV}^{-2}$ agree well with the lattice QCD results for $\chi_6^B$ and 
$\chi_8^B$ within errors. The agreement with Wuppertal-Budapest lattice results is especially 
remarkable as these results have relatively small errors. These lattice QCD results clearly 
disagree with the QMHRG model without repulsive interactions; see Fig.~\ref{fig:chi2468B} (lower panels). 
We thus conclude that, in order to apply the QMHRG model at large values of baryon number density 
to describe QCD, repulsive baryon-baryon interactions have to be taken into account. 
Here we also note that based on the lattice QCD estimates of the charm quark 
pressure~\cite{Mukherjee:2015mxc,Bazavov:2023xzm} we do not expect significant 
contribution from quark degrees of freedom to thermodynamics quantities below $T_{pc}$.
Therefore, thermodynamic estimates based on purely hadronic models will also agree with 
the lattice data up to $T_{pc}$.

In the above discussion, we assumed that both baryons and baryon resonances contribute equally 
to the mean field. This is certainly a very simplistic assumption. One may expect that not all 
baryon resonances contribute to the mean field at the same level because of their short lifetimes. 
As an extreme assumption, we can assume that only ground state baryons contribute to the repulsive 
mean field. As shown in  Appendix \ref{app:fluct}, we can describe the lattice results on $\chi_2^B$, $\chi_4^B$, 
$\chi_6^B$, and $\chi_8^B$ also in this case if the parameter $K$ is increased from $K=33~{\rm GeV}^{-2}$ 
to $K=100~{\rm GeV}^{-2}$. We also calculated the QCD pressure and energy density using the QMHRG model 
with repulsive mean-field interaction. We find that repulsive mean-field interaction has a significant 
effect on the pressure and energy density for $\mu_B/T \leq 2$, and its inclusion improves the agreement 
with the lattice QCD data. This is shown in the Appendix \ref{app:eos}. We will return to the 
calculations of the equations of state within the QMHRG model with repulsive interactions in 
Sec.~\ref{sec:netS0} for a strangeness neutral system.

\section{Chiral condensate in the mean-field QMHRG model}\label{secIII}
The light quark condensate at nonzero temperature and density
is defined as
\begin{equation}
\langle\bar{\psi}\psi\rangle_{l,T}=\langle\bar{\psi}\psi\rangle_{l,0} + 
\frac{\partial P}{\partial m_l}~,
\label{eq.defppbar}
\end{equation} 
where $m_l$  is the light quark mass and $P$ is the pressure of the 
thermodynamic medium described by QCD. We work with two degenerate light 
quarks, $m_u=m_d=m_l$.  In the HRG model, the derivatives of the pressure with 
respect to the light quark mass can be written as the mass derivative of 
the interacting baryonic gas pressure and the ideal mesonic gas pressure.
Using Eqs.~\ref{Eq.Pint} and \ref{Eq.nb} the derivative of the interacting 
baryonic pressure in Eq.~\ref{Eq.Pint} can be further written as
\begin{eqnarray}
\left(\frac{\partial P_{int}}{\partial m_l} \right) &=& - \sum_{i=B} 
M_i \frac{\partial M_i}{\partial m_l} \int g_i 
\frac{d^3 p}{(2 \pi)^3} \frac{f_i^b}{E_i} - \sum_{i=\bar{B}} M_i 
\frac{\partial M_i}{\partial m_l} \int g_i \frac{d^3 p}{(2 \pi)^3}  
\frac{f_i^{\bar{b}}}{E_i}.
\end{eqnarray}
Here $f_i^{b\{\bar{b}\}}$ is the Fermi-Dirac distribution function
corresponding to the baryons and antibaryons with the modified chemical potential 
$\mu_{eff}=B_i\mu_i - K n_{b\{\bar{b}\}} $. In the limit $K=0$ one obtains
the ideal gas result of Ref.~\cite{Biswas:2022vat}. The calculations of the mass derivative
of the mesonic pressure are the same as in Ref.~\cite{Biswas:2022vat}.

The nontrivial input needed for the calculation of the chiral 
condensate is the dependence of the hadron and resonance masses on the 
light quark mass $m_l$. In this work, we have estimated these mass 
derivatives following Ref.~\cite{Biswas:2022vat}. The uncertainties in 
the derivatives of the hadron masses with respect to $m_l$ have been 
estimated in Ref.~\cite{Biswas:2022vat}, and we propagate these uncertainties 
when evaluating the temperature and $\mu_B$ dependence of the chiral condensate.
As in our previous work, we use the renormalized chiral condensate defined by the 
HotQCD Collaboration~\cite{Bazavov:2011nk} as
\begin{eqnarray}
\Delta^l_R=d+ m_s r_1^4 \left[\langle\bar{\psi}\psi\rangle_{l,T}-
\langle\bar{\psi}\psi\rangle_{l,0}\right]~,
\label{Eq.relation1}
\end{eqnarray}
where the parameter $r_1$ is derived from the static quark potential~
\cite{Aubin:2004wf} and $d=r_1^4 m_s (\lim_{m_l \rightarrow 0} \langle 
\bar \psi \psi \rangle_{l,0})^R$. In the chiral limit, the light quark 
condensate has only a multiplicative renormalization factor. The superscript 
$R$ denotes the renormalized quantity. Taking into account the fact 
that $(\lim_{m_l \rightarrow 0} \langle \bar \psi \psi\rangle_{l,0})^R 
= 2 \Sigma$ and using the values of the low energy constant of $SU(2)$ 
chiral perturbation theory ($\chi$PT), $\Sigma^{1/3}=272(5)$ MeV and 
$m_s=92.2(1.0)$ MeV in the $\overline{\rm MS}$ scheme at $\mu=2$ GeV from 
the FLAG 2022 review for the 2+1 flavor case~\cite{Aoki:2021kgd}, as well 
as $r_1=0.3106$ fm~\cite{MILC:2010hzw}, we obtain the value of 
$d=0.022791$. 

We estimate the chiral crossover temperature, $T_{pc}$ as the temperature where the 
renormalized chiral condensate drops to half of its vacuum value. From the lattice
QCD calculations at $\mu_B=0$ we know that this definition of the chiral
crossover temperature agrees well with the usual definition of the chiral crossover
temperature, defined as the peak position of the chiral susceptibility~\cite{Bazavov:2011nk}.
In our previous work, we showed that, using this definition of $T_{pc}$ and
the temperature dependence of $\Delta_l^R$ within the QMHRG model, one obtains 
a value of $T_{pc}$ at $\mu_B=0$ as well as for small values of $\mu_B$ that 
is only a few MeV larger than the lattice results~\cite{Biswas:2022vat}. We expect
that the above criterion to estimate $T_{pc}$ should work at larger values of $\mu_B$ 
as long as we are far away from a true phase transition. 
Therefore, in this work, we assume that the chiral transition remains a crossover
transition also for large values of $\mu_B$ and estimate the chiral  crossover
temperature using the QMHRG model and the above considerations. This is reasonable 
since there is no definite indication from lattice QCD for a true phase transition 
at large $\mu_B$, and based on symmetry arguments the transition could be a crossover 
even for very large values of $\mu_B$~\cite{Schafer:1998ef}.

\section{Results on the chiral crossover line}\label{secIV}
\subsection{Renormalized chiral condensate in the mean-field formalism:}
\begin{figure}
\subfloat{\includegraphics[width=8.6cm]{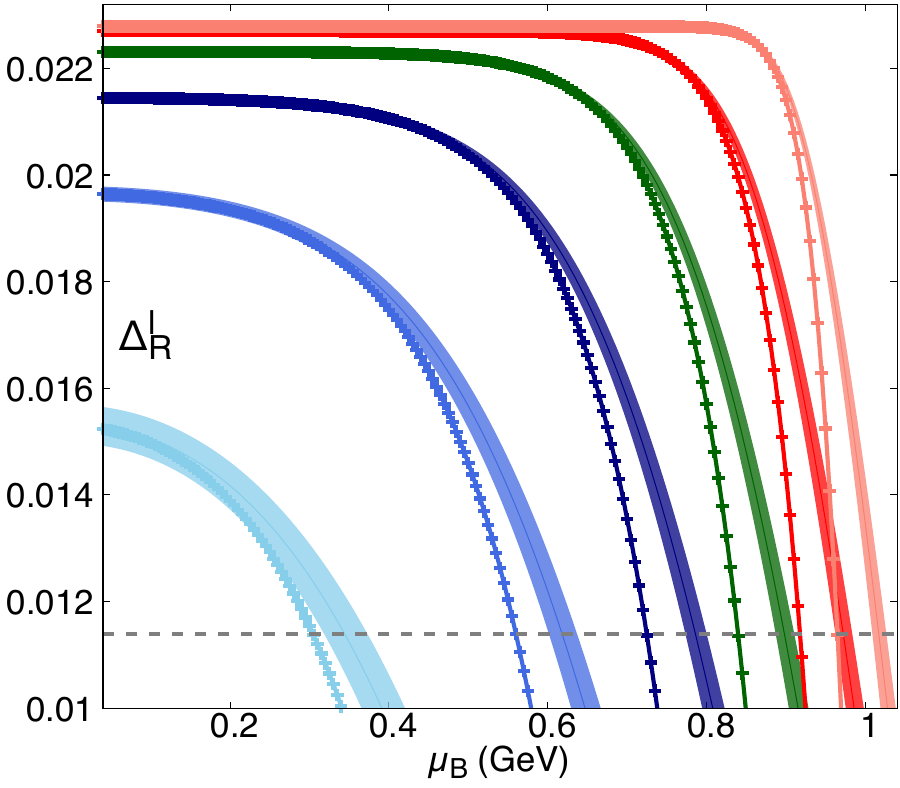}}
\subfloat{\includegraphics[width=8.6cm]{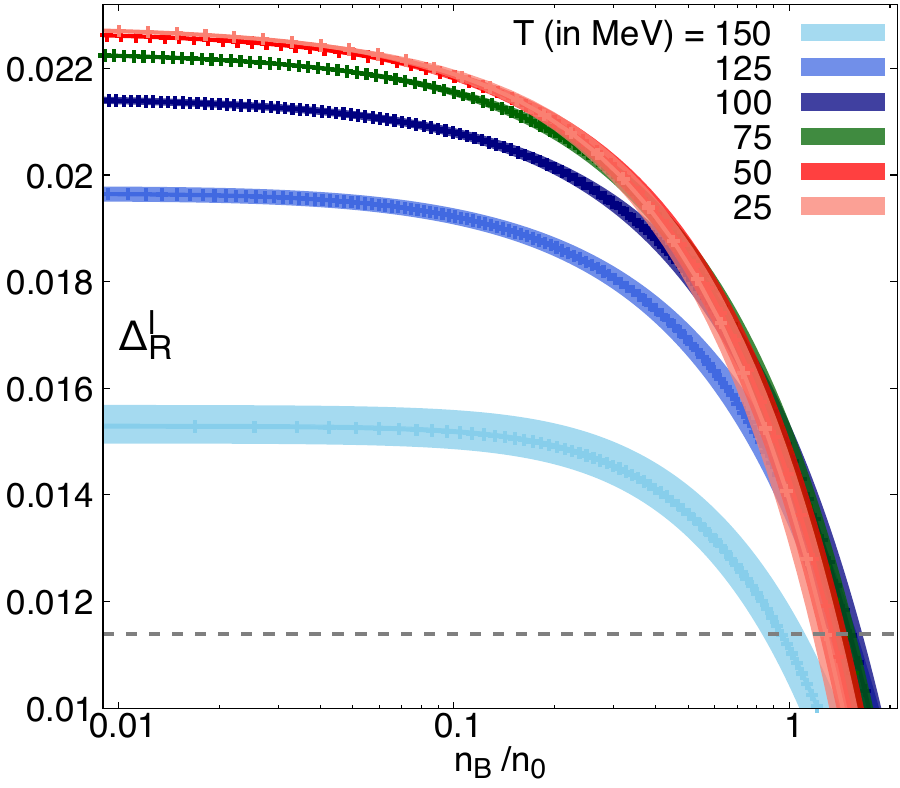}}
\caption{Left: The $\mu_B$ dependence of the renormalized chiral 
condensate at different temperatures $T=25$-$150$ MeV. The bands are 
results from the QMHRG model with repulsive mean-field interaction and 
the lines with points are the results from the ideal QMHRG model. 
Right: The variation of the $\Delta^l_R$ with net-baryon number
density normalized by the nuclear saturation density, $n_0 = 0.16~
{\text{fm}}^{-3}$, for the same range of temperatures. The widths of 
the bands represent the uncertainty of the chiral condensate due to 
the uncertainties in the quark mass derivatives of the hadron masses; 
see text for details.
}
\label{fig:DeltalR}
\end{figure}
One of the goals of this study is to study the dependence of the 
renormalized chiral condensate on temperature and baryon chemical potentials
for a large range of $\mu_B$, including values not accessible in lattice QCD
calculations. With the already estimated value of the mean-field parameter 
$K=33~\text{GeV}^{-2}$, we have evaluated $\Delta^l_R$ as a function of $\mu_B$ 
for different temperatures, shown in Fig.~\ref{fig:DeltalR}. We have also shown the 
$\Delta^l_R$ calculated in the noninteracting QMHRG model for comparison. 
In the figure, the uncertainty in the values of renormalized chiral condensate
obtained in the QMHRG model with repulsive mean-field interactions due to the 
uncertainties of the quark mass derivatives is represented as the width of the lines. 
For the noninteracting QMHRG model, we do not show the corresponding uncertainties 
for better visibility. For temperatures $T<100~$ MeV, we observe a prominent plateau 
in the value of $\Delta^l_R$ for a range of baryon chemical potentials, the extent of 
which reduces with increasing temperature. Furthermore, we observe that the renormalized 
condensate falls faster with increasing $\mu_B$ from the plateau region for the noninteracting 
QMHRG model compared to the QMHRG model with repulsive mean-field interactions. Thus the repulsive 
mean-field pushes the transition point toward larger values of $\mu_B$. As stated above we 
use the condition of the chiral condensate dropping to half of its vacuum values to estimate 
the crossover point. Thus using the values of $\mu_B$ where $\Delta_l^R$ is at half of 
its vacuum value for each temperature we obtain the estimate of crossover line as a function 
of baryon chemical potential, $T_{pc}(\mu_B)$.

We are also interested in estimating the chiral crossover temperature as a function of the 
net baryon density, $n_B$. Therefore, we have also studied the dependence of $\Delta_l^R$ 
on $n_B$. The right panel of Fig.~\ref{fig:DeltalR} shows the change in $\Delta^l_R$ 
at different temperatures as a function of net baryon number density $n_B$ scaled by 
the nuclear saturation density $n_0 = 0.16~{\text{fm}}^{-3}$. At a fixed temperature, 
$\Delta^l_R$ decreases as the net baryon density increases because baryons significantly 
contribute to the reduction of the chiral condensate. From this figure, we observe that 
the density corresponding to the chiral crossover transition for $T=100$ MeV slightly exceeds 
that for $T=25$ MeV, for example. This seemingly contradicts the usual expectation that 
$n_B$ decreases as $T$ increases along the chiral crossover line. This is an artifact of the 
hadron resonance gas model at very low temperatures, as discussed in Sec.~\ref{sec:nbvsTpc}. 

\subsection{Curvature of the crossover line in the repulsive mean-field 
QMHRG model}
\begin{figure}
\begin{center}
{\includegraphics[width=8.6cm]{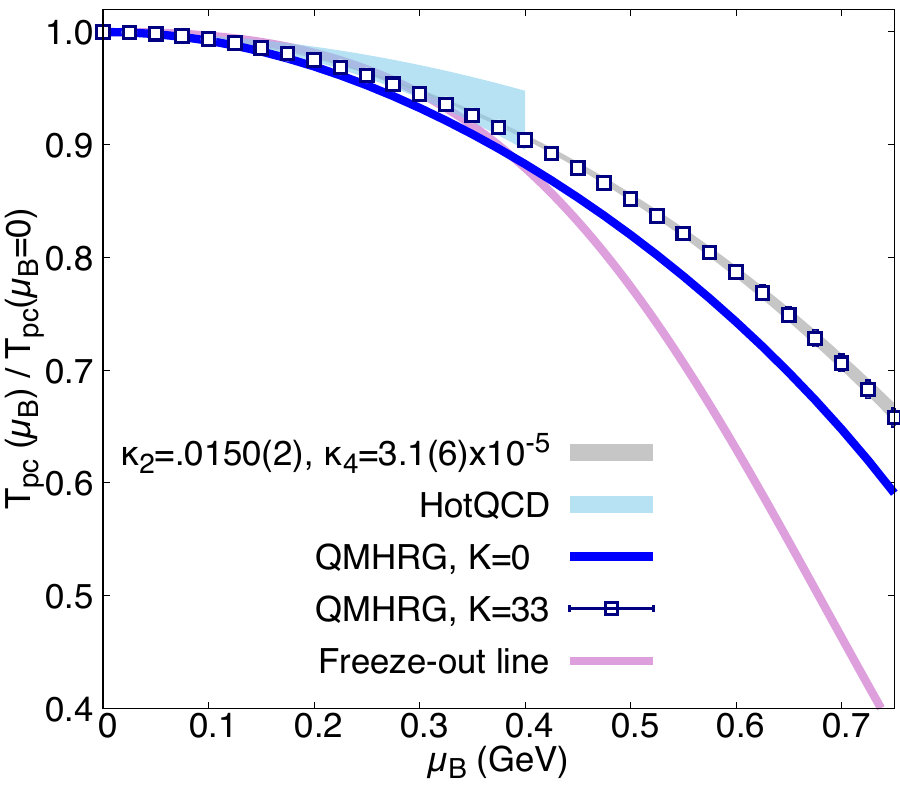}}
{\includegraphics[width=8.6cm]{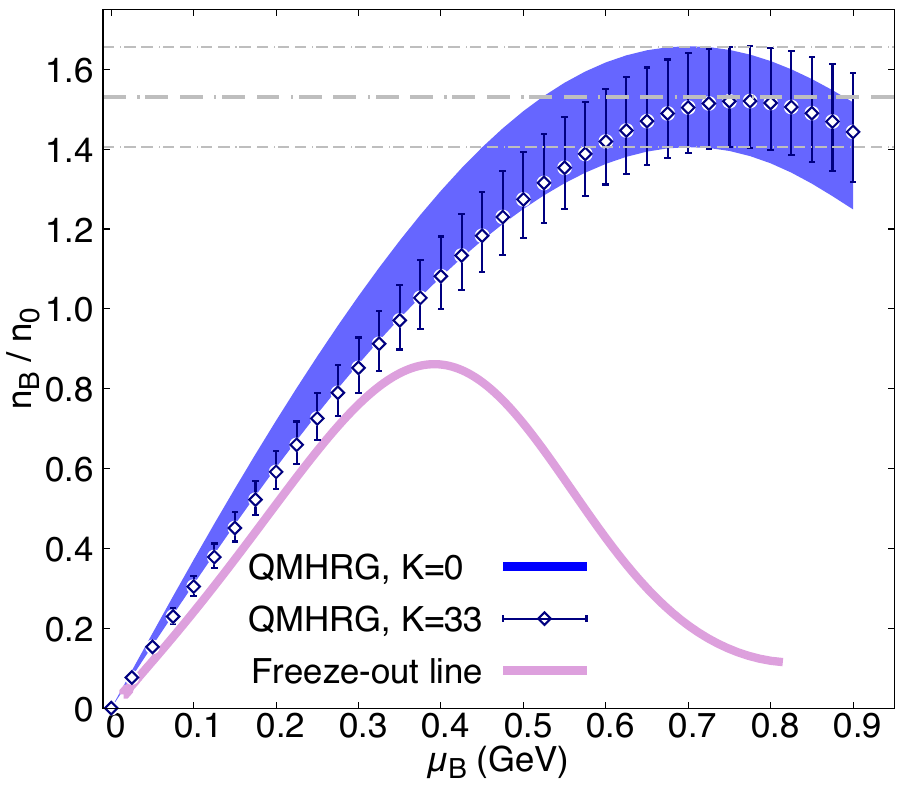}}
\end{center}
\caption{Left: The pseudocritical line calculated from the 
QMHRG model is compared with the lattice QCD results from the 
HotQCD Collaboration (light blue) from Ref.~\cite{HotQCD:2018pds}. 
The ideal gas ($K=0$) results are shown as a line 
and results including the mean-field repulsion ($K=33$) are shown 
as open symbols. These are compared with the freeze-out line (magenta) 
using the parametrization in Ref.~\cite{Andronic:2017pug}. The gray line 
represents the fit performed by us to obtain the pseudocritical line; 
see text for details. Right:  The net-baryon density in units of $n_0=0.16~{\rm fm}^{-3}$ 
is shown as function of $\mu_B$ along the pseudocritical line. 
The horizontal gray band represents the variance in the $n_B/n_0$, which is 
due to the variation of $T_{pc}$ arising from the uncertainties in 
the sigma terms. We also show the variation of the net baryon density 
along the freeze-out line. The color scheme is the same as in the left panel.}
\label{fig:phasediamu}
\end{figure}
As discussed in the previous subsection, the temperature at which the 
observable $\Delta^l_R$ falls to half its zero-temperature value is 
used as an estimate of the pseudocritical temperature $T_{pc}$~\cite{Biswas:2022vat} 
for each value of $\mu_B$. The extracted pseudo-critical temperature as a function 
of $\mu_B$, $T_{pc}(\mu_B)$, is shown as a function of $\mu_B$ in the left panel 
of Fig.~\ref{fig:phasediamu} using QMHRG model for $K=0$ and $K=33~{\rm GeV}^{-2}$.
The uncertainty in the value of $\Delta_l^R$ results in an uncertainty in the value 
of the crossover temperature, which, however, is quite small, about the size of the 
symbols in Fig. \ref{fig:phasediamu} or even smaller. In the same figure, the vertical 
axis is scaled by the estimated pseudocritical temperatures at zero baryon density, 
which are $T_{pc}(\mu_B=0)=161.2(1.7)~$MeV for the ideal case~\cite{Biswas:2022vat} 
and $T_{pc}(\mu_B=0)=161.5(1.6)~$ MeV for the repulsive QMHRG model with 
the mean-field coefficient $K=33~\text{GeV}^{-2}$. 
Our analysis is restricted to $\mu_B\leq 750$ MeV for reasons explained below.
As one can see from the figure, the effects of repulsive baryon interactions start 
to have an impact on the value of $T_{pc}$ for $\mu_B>300$ MeV, pushing the chiral crossover 
temperature toward larger values and leading to a better agreement with the lattice 
QCD results on the curvature of the pseudocritical line. On the other hand for $\mu_B < 300$ 
MeV the effects of the repulsive mean-fields are small, and good agreement for the curvature 
of the pseudocritical line is obtained using QMHRG model with or without the repulsive 
mean fields included.

In Fig.~\ref{fig:phasediamu} we also show the recent parametrization of the chemical 
freeze-out line in heavy ion collisions from Ref.~\cite{Andronic:2017pug}. The freeze-out 
line is normalized by the freeze-out temperature $T_f=158$ MeV at $\mu_B=0$~\cite{Andronic:2017pug}. 
For $\mu_B<300$ MeV the freeze-out line agrees well with the pseudocritical line.
However, we observe significant differences between the pseudocritical line and the
chemical freeze-out line for $\mu_B>400$ MeV, which become larger as the value of the 
baryon chemical potential increases. Namely, the chiral crossover temperature is always 
larger than the chemical freeze-out temperature. For $\mu_B=400$ MeV, the chiral crossover 
temperature is more than $8$ MeV larger than  the freeze-out temperature while at $\mu_B=750$ 
MeV it is more than $20$ MeV larger than the freeze-out temperature. Thus, the systems produced
in heavy ion collisions in the fixed target mode of the beam energy scan (BES) program at the 
Relativistic Heavy Ion Collider (RHIC) as well 
as in heavy ion collisions at GSI are expected to undergo a significant nonequilibrium 
chemical evolution in the hadronic phase unlike the matter produced in heavy ion collisions 
at higher center-of-mass energies, like the heavy ion collisions at the Large Hadron Collider (LHC) and at RHIC in the collider mode.

We next performed a fit to the obtained values of $T_{pc}(\mu_B)$ with the following widely used ansatz:
\begin{equation}
\label{eq:curvature}
 \frac{T_{pc}(\mu_B)}{T_{pc}(0)}=1-\kappa_2 \left(\frac{\mu_B}{T_{pc} (0)}\right)^2 - 
\kappa_4 \left(\frac{\mu_B}{T_{pc} (0)}\right)^4~.
\end{equation}
It turns out that a good fit can be obtained by this ansatz for $\mu_B \le 750$ MeV 
which results in the following  values of the curvature coefficients: 
$\kappa_2=0.0150(2)$ and $\kappa_4=3.1(6)\times 10^{-5}$. The value of $\kappa_2$ 
is consistent within errors with the two most recent continuum estimates from lattice 
QCD studies with physical masses, $\kappa_2=0.012(4)$~\cite{HotQCD:2018pds}
and $\kappa_2=0.0153(18)$~\cite{Borsanyi:2020fev}, as well as with the earlier 
lattice QCD estimates~\cite{Bonati:2018nut,Bellwied:2015rza,Bonati:2015bha}.
For the first time, we could estimate a nontrivial value of $\kappa_4$ 
which is distinctly different from zero given the estimated errors. Lattice
QCD calculations report a $\kappa_4$ which is compatible with zero within
errors~\cite{HotQCD:2018pds, Borsanyi:2020fev}. The above estimate for the 
leading curvature coefficient is somewhat smaller than the one obtained from 
the relativistic nuclear mean-field model calculation, while $\kappa_4$ has 
the opposite sign~\cite{Abhishek:2023uha}. Allowing for a nonzero value of $\kappa_6$ 
does not improve the quality of the fit, and only leads to much larger errors in $\kappa_2$ 
and $\kappa_4$. Therefore, it is fair to say that, within our accuracy, $\kappa_6$ is 
consistent with zero.


\subsection{Net-baryon density, energy density, and particle
composition along  the pseudocritical line}
\label{sec:nbvsTpc}
The right panel of Fig.~\ref{fig:phasediamu} shows the estimated values of 
net-baryon density normalized by the nuclear saturation density $n_0$ 
as function of $\mu_B$ along the pseudocritical line in the $T-\mu_B$ plane.
The baryon density increases monotonically with $\mu_B$ as expected, but 
for sufficiently large $\mu_B$ it reaches a maximum and then decreases with 
a further increase in $\mu_B$. This behavior is clearly unphysical and therefore, 
the applicability of the HRG model is restricted for large values of $\mu_B$.
The maximum in $n_B$ of $1.53(13)n_0$  is reached at $\mu_B=700$ MeV when $K=0$ 
and at $\mu_B=750$ MeV for $K=33~{\rm GeV}^{-2}$. Thus the inclusion of repulsive 
mean-field interaction in the QMHRG model calculation helps to extend the model to slightly 
larger values of $\mu_B$. However, our QMHRG model with a single repulsive mean field 
is too simplistic and is not valid for very large values of $\mu_B$. The large value 
of $n_B$ obtained at relatively small values of $\mu_B$ is due to the contribution 
from baryon resonances. Since the chiral crossover temperature decreases with increasing 
$\mu_B$, the contribution from these states will eventually decrease and the large value 
of the net-baryon density near the crossover point should come from nucleons and possibly 
their attractive interactions. But the simple HRG model cannot describe this switchover. 
Thus for temperatures close to the chiral crossover temperatures, the QMHRG model is only 
applicable for $\mu_B<750$ MeV.

The nonmonotonic variation of $n_B$ with $\mu_B$ is also observed in the 
Nambu-Jona-Lasinio model~\cite{Schwarz:1999dj}, where the mass-gap 
equation determines the phase transition line, resulting in a qualitatively 
similar trend. A similar maximum of the net-baryon density as a function of 
$\mu_B$ was earlier observed in a HRG motivated study along the freeze-out 
boundary in Ref.~\cite{Randrup:2006nr}, which was determined from the hadron 
yields at different collision energies. 
\begin{figure}[ht]
\subfloat{\includegraphics[width=8.6cm]{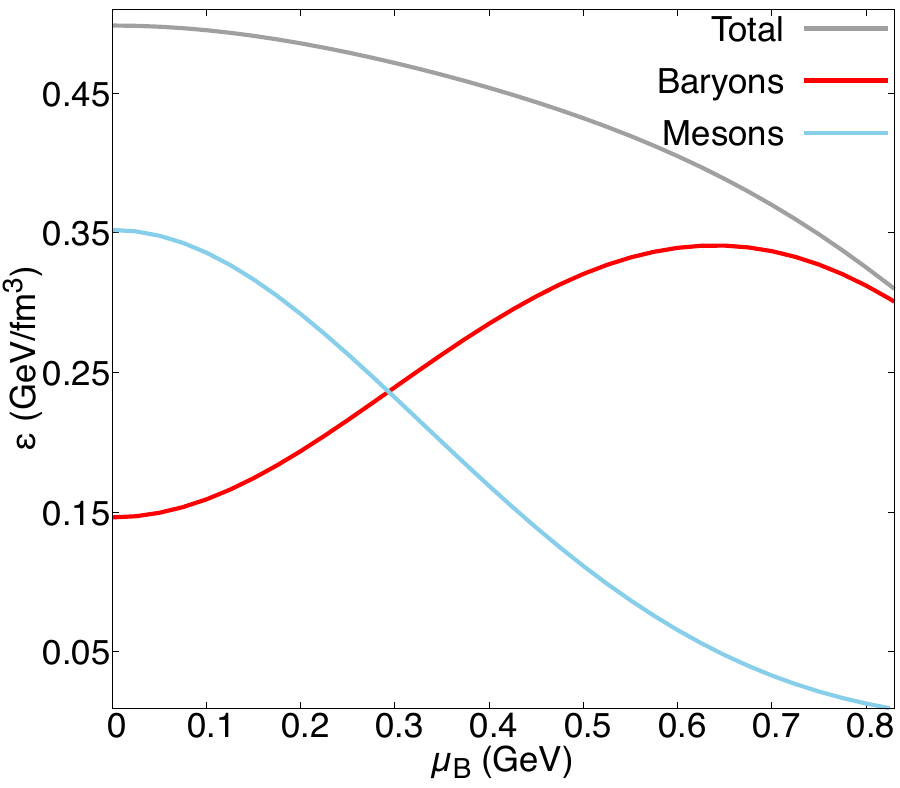}}
\subfloat{\includegraphics[width=8.6cm]{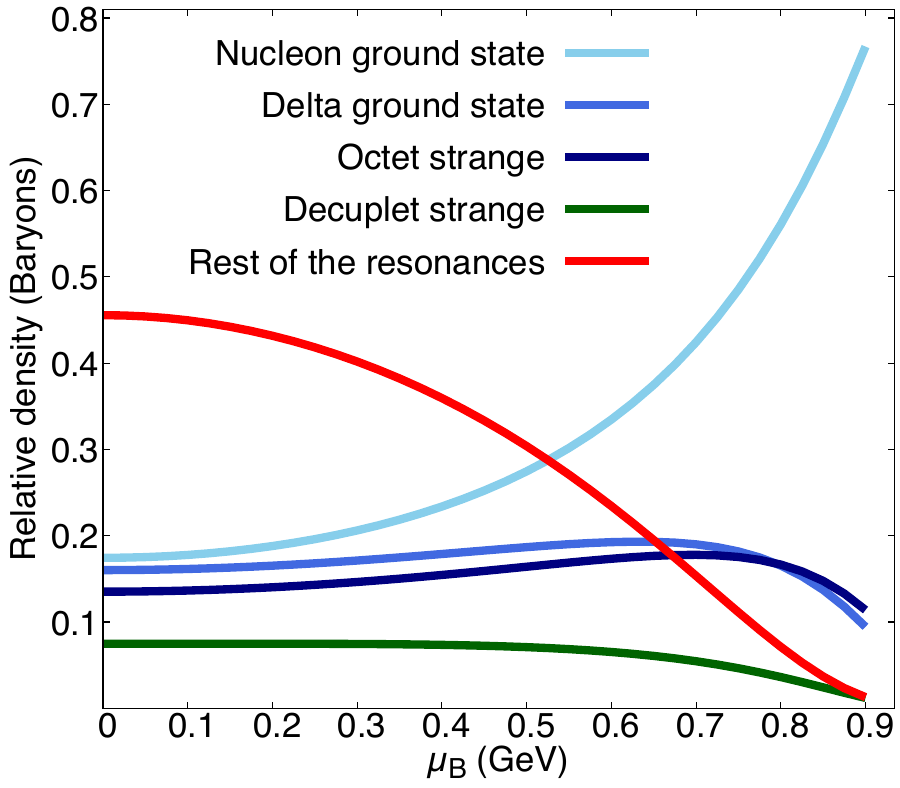}}
\caption{ Left: Relative contributions to the energy density from mesons (blue)
and baryons (red) along the pseudo-critical line compared to the total energy density 
(gray) calculated in the mean-field QMHRG model. Right: The relative number densities of 
various baryon species shown along the pseudo-critical line.}
\label{fig:Eosfinitemu}
\end{figure}

It is interesting to study the variation of energy density along the pseudocritical 
line, which is shown in the left panel of Fig.~\ref{fig:Eosfinitemu}. One can 
see that the energy density along the pseudocritical line slightly decreases 
with increasing $\mu_B$. The relative contribution to the energy 
density coming from the baryon sector increases till $\mu_B=650$ MeV as the baryon 
chemical potential increases, resulting in a shift from a meson-dominated 
to a baryon-dominated scenario, beyond $\mu_B=350$ MeV. A comparable 
phenomenon can be observed in heavy-ion collisions, where a transition 
from meson to baryon dominance during freeze-out takes place at
lower energies~\cite{Cleymans:2004hj}, where the net-baryon density is larger.

Furthermore, we would like to understand the relative abundances of different
baryon species with increasing $\mu_B$. In the right panel of Fig.~\ref{fig:Eosfinitemu} 
we have shown the relative contribution of different baryons and baryon resonances to the 
net-baryon density as a function of $\mu_B$ along the pseudocritical line. 
As the baryon chemical potential increases, the relative contribution of the nucleons
also increases. The relative contributions of the lowest lying strange baryons and of the 
lowest $\Delta$ resonances do not change significantly with increasing $\mu_B$, either. 
While the net contribution from higher lying resonances decreases with increasing $\mu_B$, 
this contribution remains significant up to $\mu=750$ MeV. Even for the largest value of 
$\mu_B=750$ MeV, where our HRG model is applicable, nucleons contribute less than 50\% to 
the total net-baryon density. Thus the contribution of the baryon resonances remains
important even at large values of $\mu_B$, and this contribution is responsible
for the relatively large values of $n_B$. At the same time, the treatment of
the repulsive interactions for baryon resonances is quite simplistic.
In future, a more refined QMHRG mean-field model, which treats the repulsive interactions 
differently in the nucleon, strange baryon, and baryon resonance sectors will be needed 
to extend the reach of the QMHRG model.

\subsection{Strangeness neutrality and the chiral pseudo-critical line}
\label{sec:netS0}
\begin{figure}[ht]
\subfloat{\includegraphics[width=8.6cm]{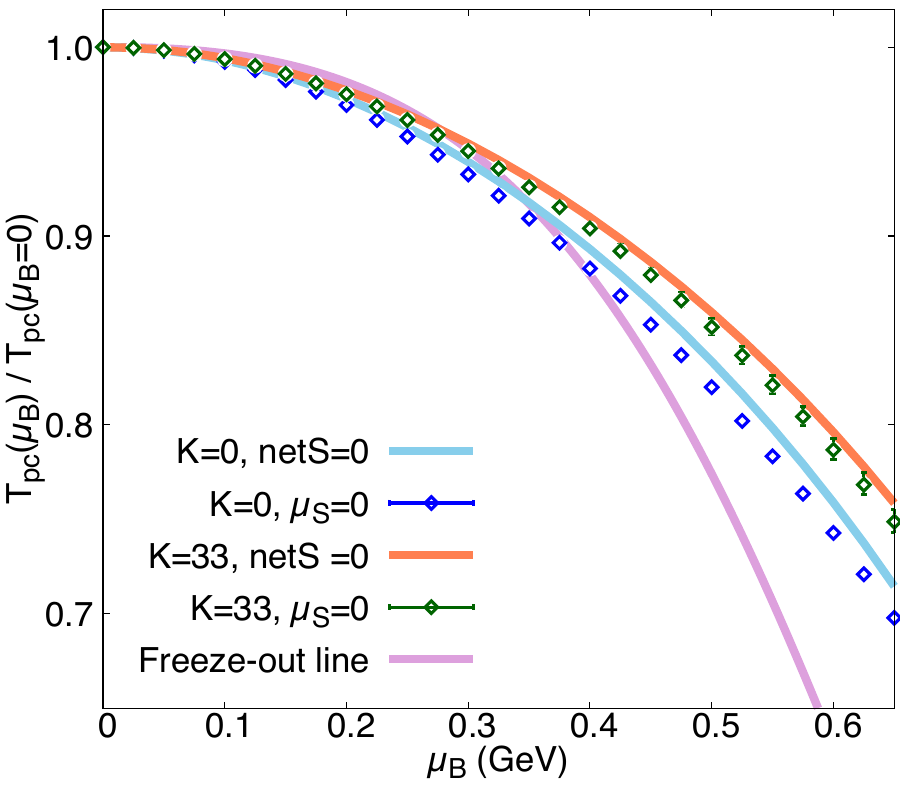}}
\subfloat{\includegraphics[width=8.6cm]{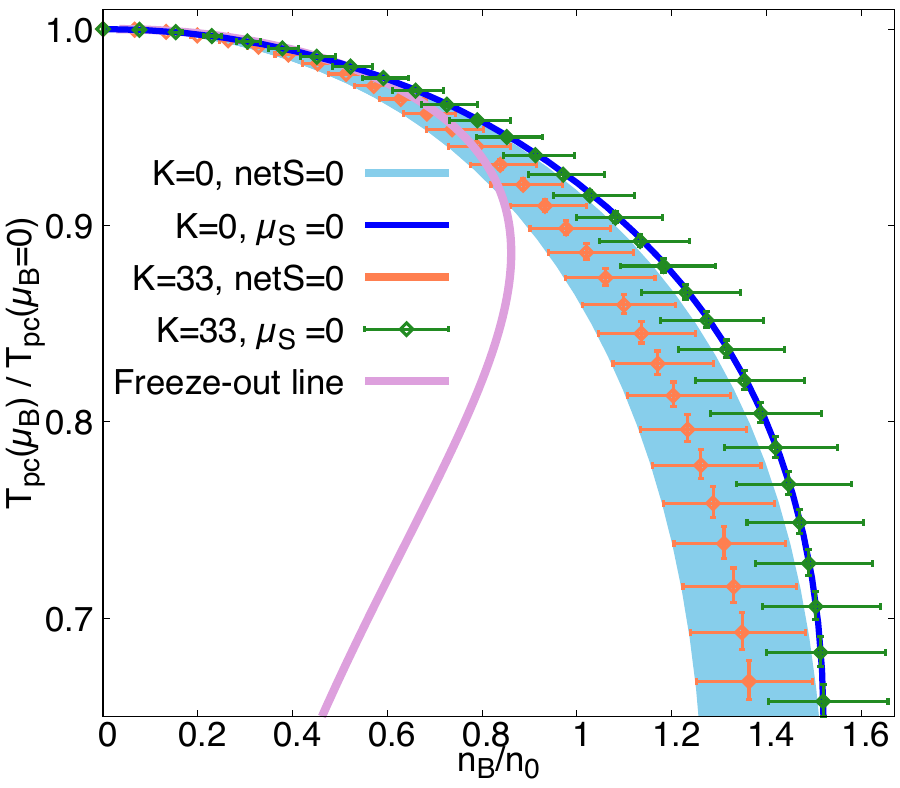}}
\caption{ The chiral crossover temperature at finite density $\mu_B$ normalized 
by its corresponding value at $\mu_B=0$ is shown as function of $\mu_B$ (left)
and as a function of $n_B$ (right) for $\mu_S=0$ and under strangeness
neutrality condition, $n_S=0$. Ideal QMHRG model results are shown in blue. Coral and 
green represents mean-field QMHRG results for $n_S=0$ and $\mu_S=0$ respectively.
}
\label{fig:Tmu_netS}
\end{figure}

The system created in heavy ion collisions has zero netstrangeness, since
the incoming nuclei do not contain strange particles. Therefore, it is important
to estimate the pseudocritical line for the conditions realized in heavy-ion experiments,
namely for zero net strangeness, $n_S=0$. At nonzero values of $\mu_B$, one needs to
tune the strangeness chemical potential $\mu_S$ to achieve zero netstrangeness.
We performed this tuning to estimate the crossover temperature and the equation of
state within our QMHRG model. The values of $\mu_S$ needed to achieve strangeness
neutrality are shown in Appendix~\ref{app:muS}.

In Fig.~\ref{fig:Tmu_netS} (left) we show the crossover temperature
in the case of $n_S=0$ as a function of the baryon chemical potential, while
in Fig.~\ref{fig:Tmu_netS} (right) we show the crossover temperature as a function 
of the net-baryon number. For a comparison, we also show our results for $T_{pc}$
for the unconstrained, $\mu_S=0$ case in the same figure. We observe that imposing 
strangeness neutrality pushes the crossover temperature to a slightly larger 
values compared to the case of $\mu_S=0$ discussed in the previous subsections, 
whether or not one includes the effect of the repulsive mean field.
The effect of imposing strangeness neutrality on the crossover temperature 
is considerably smaller than the effect of the repulsive interactions.
Interestingly enough, the effects of the repulsive interactions 
are not visible on $T_{pc}$ when plotted as a function of the net-baryon density.
This can seen from Fig.~\ref{fig:Tmu_netS} (right). The clear difference between 
the pseudocritical line and the freeze-out line is also apparent for the zero net-strangeness 
case, cf. Fig.~\ref{fig:Tmu_netS}, especially when the results are shown 
in the $T-n_B$ plane. 

The maximum net-baryon number density that can be reached within our model is slightly reduced
compared to the unconstrained, $\mu_S=0$ case to about $1.4 n_0$ compared to $1.53 n_0$
as obtained in the previous subsection. This is due to the fact that a positive nonzero
value of $\mu_S$ required by the condition $n_S=0$ reduces the contribution of strange
baryons to the partition function.

\section{Speed of sound along the isentropic lines}\label{secV}
\begin{figure}[ht]
\subfloat{\includegraphics[width=8.6cm]{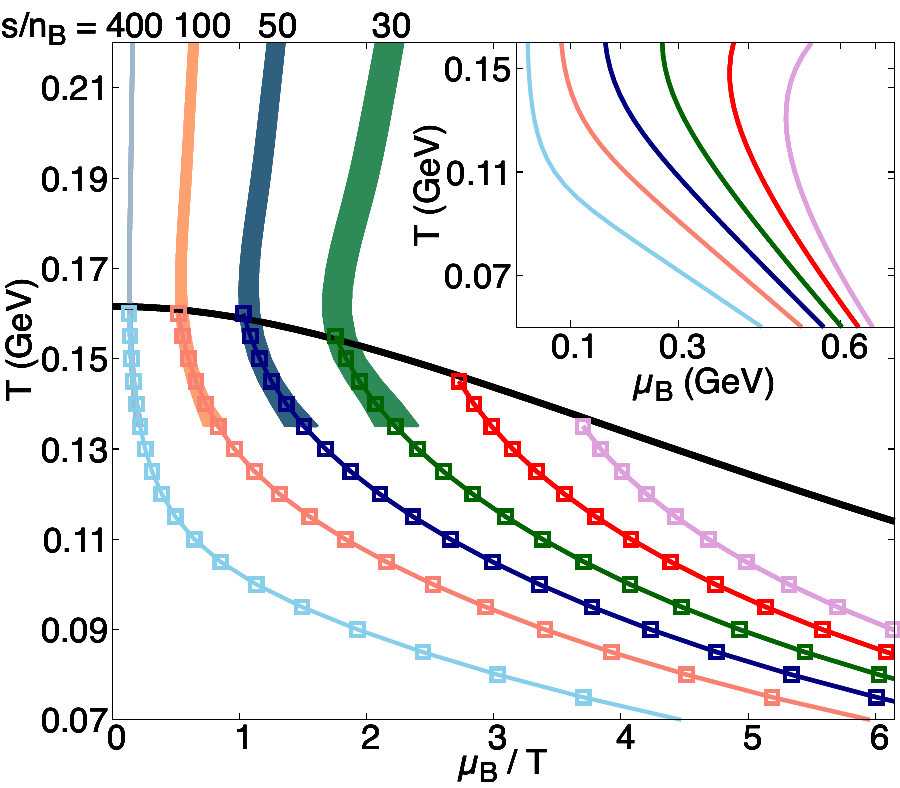}}
\subfloat{\includegraphics[width=8.6cm]{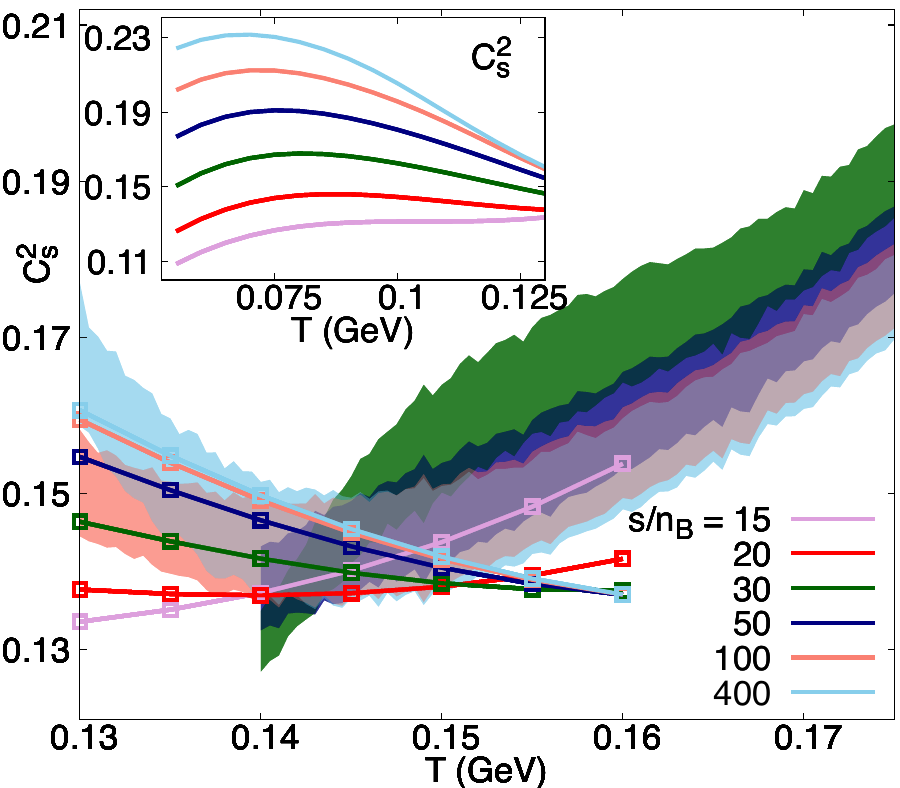}}
\caption{ Left: The isentropes for $s/n_B= 15, 
20, 30, 50, 100$, and $400$ respectively are shown in the $T-\mu_B$ plane. 
Right: The speed of sound as a function of the temperature is shown along these 
isentropes. The bands are the lattice QCD data from Ref.~\cite{Bollweg:2022fqq} and 
the points are the QMHRG model calculations with repulsive mean field. The isentropes 
and the speed of sound in the low temperature region $60<T<150$ MeV are shown in the 
insets of the left and the right figures respectively. The black line 
represents the pseudocritical line evaluated within our mean-field QMHRG model.} 
\label{fig:sbynBcs2}
\end{figure}
In heavy-ion experiments, the chemical freeze-out surface is typically
defined as a surface of constant entropy per baryon density, denoted as 
$s/n_B$ \cite{Cleymans:1998fq}. Hence it is interesting to study the 
evolution of the QCD matter formed in heavy-ion collisions along the 
lines of constant $s/n_B$. The evolution of this matter is subject to 
several constraints, including strangeness neutrality ($n_S=0$) and a fixed 
ratio of net electric charge ($n_Q$) to net-baryon number ($n_B$) density 
due to charge conservation. We study a particular case of strangeness-neutral 
system with isospin symmetry which is realized when $n_Q/n_B=0.5$, 
corresponding to a nonzero $\mu_B$ and $\mu_S$ but $\mu_Q=0$. It has 
been shown that the differences in bulk thermodynamic observables that arise 
from deviations in $n_Q/n_B$ from the isospin-symmetric value of $0.5$ are 
minimal~\cite{Bollweg:2022fqq}. We have examined the isospin-symmetric 
scenario for a range of entropy per baryon density values, $s/n_B=15, 20, 30, 50, 
100,$ and $400$. These values effectively cover the range of center-of-mass energies 
explored in the RHIC BES program in the collider mode~\cite{Bollweg:2022fqq}. 
The left panel of Fig.~\ref{fig:sbynBcs2} illustrates the isentropic trajectories 
for different values of $s/n_B$ and compares to the lattice QCD results 
available up to $\mu_B/T=2.5$ from Ref.~\cite{Bollweg:2022fqq}. We have also 
included the pseudocritical line estimated by us in the same plot for 
comparison. Our estimates obtained within the mean-field repulsive QMHRG 
model show a good agreement with the lattice results at temperatures 
below the pseudo-critical temperatures. For completeness, we have also 
included the isentropic trajectories for $s/n_B= 15$ and $s/n_B=20$ 
calculated within the repulsive QMHRG model. These trajectories are beyond 
the reach of lattice QCD calculations and hold significance for lower collision 
energies of the RHIC BES program in the fixed target mode~\cite{Motornenko:2019arp}.

In the right panel of Fig.~\ref{fig:sbynBcs2}, we have shown the speed of 
sound calculated within the mean-field repulsive QMHRG model calculated 
along these constant $s/n_B$ trajectories. The speed of sound is defined 
as~\cite{Bollweg:2022fqq} 
\begin{eqnarray}
    c_s^2=\left(\frac{\partial P}{\partial \epsilon}\right)_{s/n_B,n_S, n_B/n_Q} 
    =\frac{\left(\partial P / \partial T \right)_{s/n_B,n_S, n_B/n_Q}}{\left(\partial \epsilon / \partial T \right)_{s/n_B,n_S, n_B/n_Q}}
\end{eqnarray}
Here, all the derivatives are taken along the lines of constant 
$n_S=0, n_B/n_Q=0.5$ for different values of $s/n_B$. We show the speed 
of sound for our QMHRG model and compare it with the latest lattice QCD 
results from Ref.~\cite{Bollweg:2022fqq}. The $c_s^2$ calculated within the 
mean-field QMHRG model decreases as the temperature is increased towards 
$T_{pc}$ for $s/n_B> 30$.  The QMHRG results with repulsive mean field 
for $c_s^2$ are in agreement with the lattice QCD results 
for $T<150$ MeV. Beyond this temperature, the lattice QCD results show 
an increasing trend with the temperature, which cannot be reproduced within a 
QMHRG model, even when the effect of the repulsive interactions are included.
We also observe that in the temperature range $100$ $< T <  150$ MeV 
the speed of sound decreases with increasing temperature both in the lattice
QCD calculations and in the QMHRG model.  This leads to a minimum of the speed 
of sound in the range of temperatures $140$-$150$ MeV. From the inset 
of Fig.~\ref{fig:sbynBcs2} (right) we also observe that the speed of sound has 
a maximum in the low temperature region for $s/n_B\ge 30$. 

The behavior of the speed of sound for $s/n_B=20$ and $s/n_B=15$, currently 
inaccessible through lattice QCD techniques, is somewhat different. From 
Fig.~\ref{fig:sbynBcs2} (right) we observe that for $s/n_B=20$ the speed of 
sound has a very shallow minimum around $T=145$ MeV, while for $s/n_B=15$ it 
is always monotonically increasing with increasing temperatures. This implies 
that the maximum in $c_s^2$, evident from the plot at low temperatures $T<75$ MeV 
and small baryon densities, should disappear when baryon densities are increased. 
We thus do not observe a softening of the equation of state near the chiral 
crossover transition at large values of baryon densities corresponding to 
$s/n_B \approx 15$ within this mean-field treatment of repulsive baryon interactions.

\section{Summary}\label{secVI}

In this paper, we have studied the QCD equation of state and the chiral crossover 
at non-zero net baryon density using the QMHRG model with repulsive mean-field interaction.
We have shown that this model can describe the lattice results on the fluctuations
of net baryon number up to the eighth order in the vicinity of the chiral
crossover temperature if the parameter characterizing the strength
of the mean-field repulsive interactions is properly chosen. This is contrary to
the usual QMHRG which can describe the higher-order net baryon number fluctuations only 
for considerably lower temperatures. 

We extended our previous calculation of the chiral crossover temperature within the QMHRG model
limited to the region of small baryon chemical potential \cite{Biswas:2022vat} to
much larger values of the baryon chemical potential. We showed that for $\mu_B>400$ MeV
there is a significant effect of the repulsive mean field on the value of the chiral crossover
temperature. We found that the $\mu_B$ dependence 
of the chiral crossover temperature can be very well parametrized by the following form:
$T_{pc}(\mu_B)/T_{pc}(0)=1-\kappa_2 (\mu_B/T_{pc}(0))^2-\kappa_4 (\mu_B/T_{pc}(0))^4$
with $\kappa_2=0.0150(2)$ and $\kappa_4=3.1(6) \times 10^{-5}$.
The value of the leading curvature coefficient, $\kappa_2$, agrees well with the lattice QCD results 
within errors. The smallness of $\kappa_4$ is consistent with the upper bounds from lattice QCD, but for
the first time we found that the corresponding value is significantly different from zero.

To realize the relevance of this study in the context of heavy-ion collision 
experiments, we have estimated the chiral crossover line imposing the 
strangeness-neutrality condition. The separation between the freeze-out 
curve and the pseudocritical line increases toward high density, which 
implies a longer-lived interacting hadronic phase at lower collision energies. 
Furthermore, we have used the repulsive mean-field QMHRG model to calculate the 
speed of sound along various isentropic ( constant $s/n_B$) lines 
pertinent to heavy-ion 
collision experiments, constrained by the strangeness neutrality and 
$n_Q/n_B$ ratio. In the hadron phase, there is good agreement between 
our calculation and the latest available data from lattice QCD.  We have 
also predicted the isentropic trajectory and the speed of sound within this 
repulsive mean-field model for $s/n_B = 15$, where no lattice QCD data 
are available. We see that starting from this value of $s/n_B$ the previously seen
characteristic softening of the QCD equation of state near the chiral crossover regions disappears.

We pointed out that the QMHRG model with a single repulsive mean field for all baryons only works
for $\mu_B<750$ MeV. For larger values of the baryon chemical potential, a more refined mean field
approach is needed, with different mean fields for different baryon species.

\section*{Acknowledgments}
D.B. is supported by the Department Of Atomic Energy (DAE), India. P.P. is supported by the U.S. Department of Energy, Office of Science, through Contract No. DE-SC0012704. S.S. gratefully acknowledges support from the Department of Science and Technology, Government of India through a Ramanujan Fellowship when a major part of this work was done. D.B. gratefully acknowledges Jishnu Goswami for helping with the lattice data from the HotQCD Collaboration. Additionally, D.B. would like to extend thanks to A. Abhishek, H. Mishra, and N. Sarkar for useful discussions. D.B. thanks P. Parotto for his assistance with the continuum extrapolated lattice data from the WB Collaboration.
\appendix

\section{Calculation of baryon number fluctuations within mean-field HRG}
\label{app:fluct}

The fluctuations of baryon charges can be derived from the pressure using \cite{Bazavov:2020bjn}

\begin{equation}
\chi_{B}^{n}=\frac{\partial^n \left[P(\mu_B/T)/T^4\right]}{\partial (\mu_B/T)^n }
\end{equation}

In the mean-field formalism, the respective number densities for baryons 
and antibaryons in the Boltzmann approximation are written as \cite{Huovinen:2017ogf}

\begin{equation}
n_b=\sum_{i=B} \int g_i \frac{d^3 p}{(2 \pi)^3} e^{-\beta(E_i-\mu_B+K n_b)}~~~{\text{and}}~~~ 
n_{\bar{b}}=\sum_{i=\bar{B}} \int g_i \frac{d^3 p}{(2 \pi)^3}
e^{-\beta(E_i+\mu_B+K n_{\bar{b}})} .
\end{equation}

Here, $\beta$ is the inverse of temperature. The above equations are transcendental equations and one can evaluate 
the fluctuations of the net baryon number by taking the derivative with $\mu_B/T$,
\begin{eqnarray}
\frac{\partial n_b}{\partial (\beta \mu_B)}&=&\sum_{i=B} g_i \int \frac{d^3 p}{(2 \pi)^3}e^{-\beta(E_i-\mu_B+K n_b)}
\left[1 - \beta K \frac{\partial n_b}{\partial (\beta \mu_B)} \right]  = \sum_{i=B} n_i
\left[1 - \beta K \frac{\partial n_b}{\partial (\beta \mu_B)} \right]  \nonumber \\
\end{eqnarray}

By re-arranging the above equation, one can write the first derivative as 
\begin{equation}
n_b'=\frac{\partial n_b}{\partial (\beta \mu_B)}= \frac{\sum_{i=B} n_i}{\left[ 1+\beta K	\sum_{i=B} n_i	\right]}=\frac{n_b}{\left[ 1+\beta K	n_b	\right]}.
\end{equation}

Similarly, for the antibaryon sector,
\begin{equation}
n_{\bar{b}}'=\frac{\partial n_{\bar{b}}}{\partial (\beta \mu_B)}= \frac{- \sum_{i=\bar{B}} n_i}{\left[ 1+\beta K	\sum_{i=\bar{B}} n_i	\right]}=\frac{-n_{\bar{b}}}{\left[ 1+\beta K	n_{\bar{b}}	\right]}. 
\end{equation}
In the mean-field formalism, the total pressure in the baryon sector is 
written as a sum of the partial pressures of the (anti)baryons \cite{Huovinen:2017ogf}, 

\begin{eqnarray}
P=T (n_b + n_{\bar{b}}) + \frac{K}{2}(n_{b}^2 + n_{\bar{b}}^2)  \nonumber ~~ {\text{and}} ~~~
\frac{P}{T^4}=\beta^3 (n_b + n_{\bar{b}}) + \beta^4 \frac{K}{2}(n_{b}^2 + n_{\bar{b}}^2).
\end{eqnarray}
Taking the first-order derivative with respect to $\beta \mu_B$, we get
\begin{eqnarray}
\chi_{B}^{1}&=&\beta^3 (n_b' + n_{\bar{b}}') + \beta^4 K(n_b n_{b}' + n_{\bar{b}} n_{\bar{b}}') 
= \beta^3 n_b'(1+\beta K n_b  ) + \beta^3 n_{\bar{b}}'(1+\beta K n_{\bar{b}} ). \nonumber \\
&=& \beta^3 \left( n_b - n_{\bar{b}}  	\right)
\end{eqnarray}
The above equation ensures that the net baryon number is given by
the difference of the number of baryons and the number of antibaryons, and vanishes
for $\mu_B=0$.

Using the 
above expressions, it is easy to extend the calculations to higher-order 
derivatives with respect to $\mu_B/T$. 
For completeness here we list all the derivatives up to the eighth order.
\begin{eqnarray}
\chi_{B}^{1}&=&\beta^3 \left[n_b - n_{\bar{b}}  	\right],  \\ 
\chi_{B}^{2}&=&\beta^3 \left[\frac{n_b}{(1+\beta K n_b)} + \frac{n_{\bar{b}}}{(1+\beta K n_{\bar{b}})}     	\right], \nonumber\\
\chi_{B}^{3}&=&\beta^3 \left[\frac{n_b}{(1+\beta K n_b)^3} - \frac{n_{\bar{b}}}{(1+\beta K n_{\bar{b}})^3}     	\right], \nonumber\\
\chi_{B}^{4}&=&\beta^3 \left[\frac{n_b (1-2\beta K n_b)}{(1+\beta K n_b)^5} + \frac{n_{\bar{b}} (1-2\beta K n_{\bar{b}})}{(1+\beta K n_{\bar{b}})^5}     	\right], \nonumber\\
\chi_{B}^{5}&=&\beta^3 \left[\frac{n_b (1-8\beta K n_b + 6 [\beta K n_b]^2 )}{(1+\beta K n_b)^7} - \frac{n_{\bar{b}} (1-8\beta K n_{\bar{b}}+6 [\beta K n_{\bar{b}}]^2)}{(1+\beta K n_{\bar{b}})^7}     	\right], \nonumber \\
\chi_{B}^{6}&=&\beta^3 \left[\frac{n_b (1-22 \beta K n_b + 58 [\beta K n_b]^2 - 24 [\beta K n_b]^3 )}{(1+\beta K n_b)^9} + \frac{n_{\bar{b}} (1-22\beta K n_{\bar{b}}+58 [\beta K n_{\bar{b}}]^2 - 24 [\beta K n_{\bar{b}}]^3)}{(1+\beta K n_{\bar{b}})^9}     	\right], \nonumber \\
\chi_{B}^{7}&=&\beta^3 \left[\frac{n_b (1-52 \beta K n_b + 328 [\beta K n_b]^2 - 444 [\beta K n_b]^3 + 120 [\beta K n_b]^4 )}{(1+\beta K n_b)^{11}}\right] - \nonumber \\
&~& \beta^3 \left[\frac{n_{\bar{b}} (1-52\beta K n_{\bar{b}}+328 [\beta K n_{\bar{b}}]^2 - 444 [\beta K n_{\bar{b}}]^3 + 120 [\beta K n_{\bar{b}}]^4)}{(1+\beta K n_{\bar{b}})^{11}}\right],     \nonumber \\
\chi_{B}^{8}&=&\beta^3 \left[\frac{n_b (1-114 \beta K n_b + 1452 [\beta K n_b]^2 - 4400 [\beta K n_b]^3 + 3708 [\beta K n_b]^4 
- 720 [\beta K n_b]^5)}{(1+\beta K n_b)^{13}}\right] + \nonumber \\
&~& \beta^3 \left[\frac{n_{\bar{b}} (1-114\beta K n_{\bar{b}}+1452 [\beta K n_{\bar{b}}]^2 - 4400 [\beta K n_{\bar{b}}]^3 + 3708 [\beta K n_{\bar{b}}]^4 - 720 [\beta K n_{\bar{b}}]^5)}{(1+\beta K n_{\bar{b}})^{13}}\right]  \nonumber. 
\end{eqnarray}
It is obvious from the above equations that the odd fluctuations of net baryon number vanish for $\mu_B=0$.

\begin{figure}[ht]
\subfloat{\includegraphics[width=8cm]{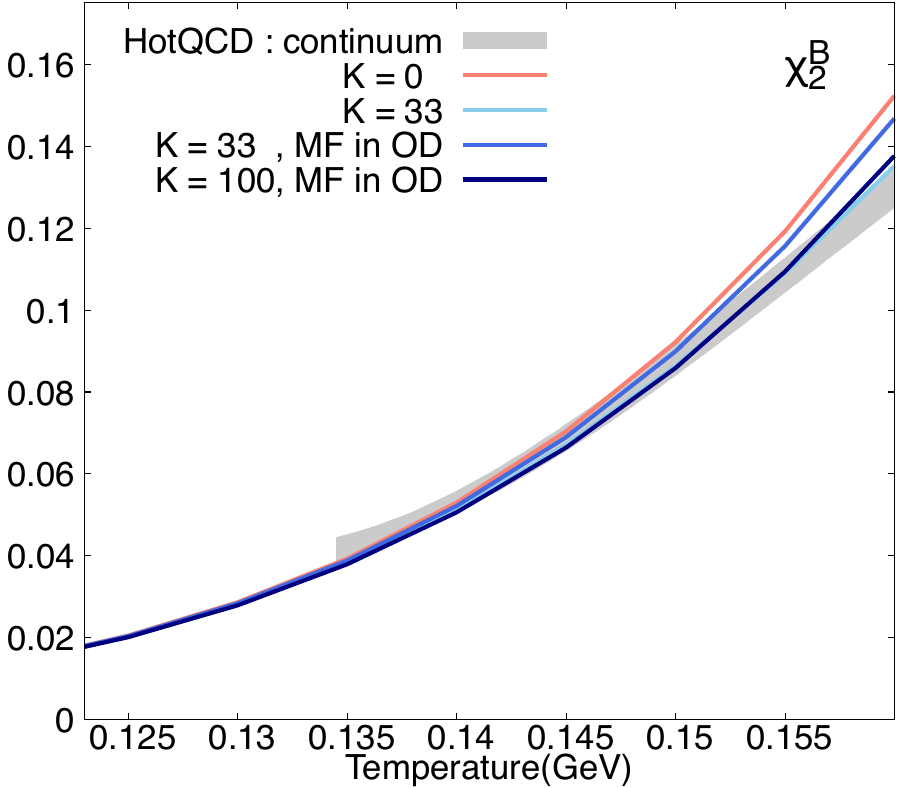}}
\subfloat{\includegraphics[width=8cm]{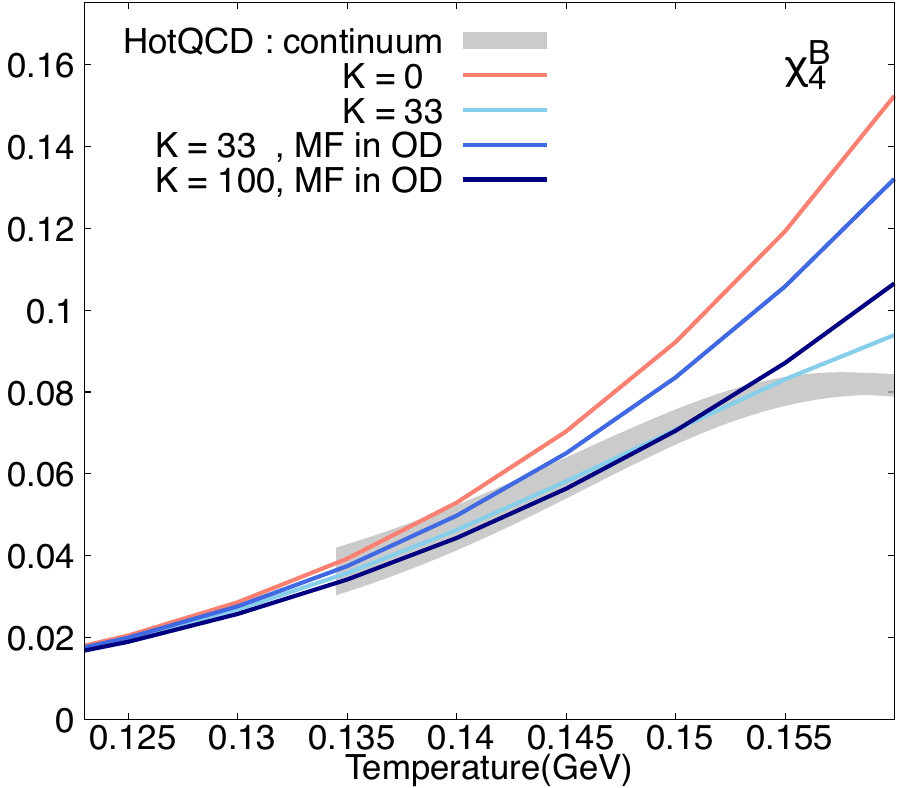}}\\
\subfloat{\includegraphics[width=8cm]{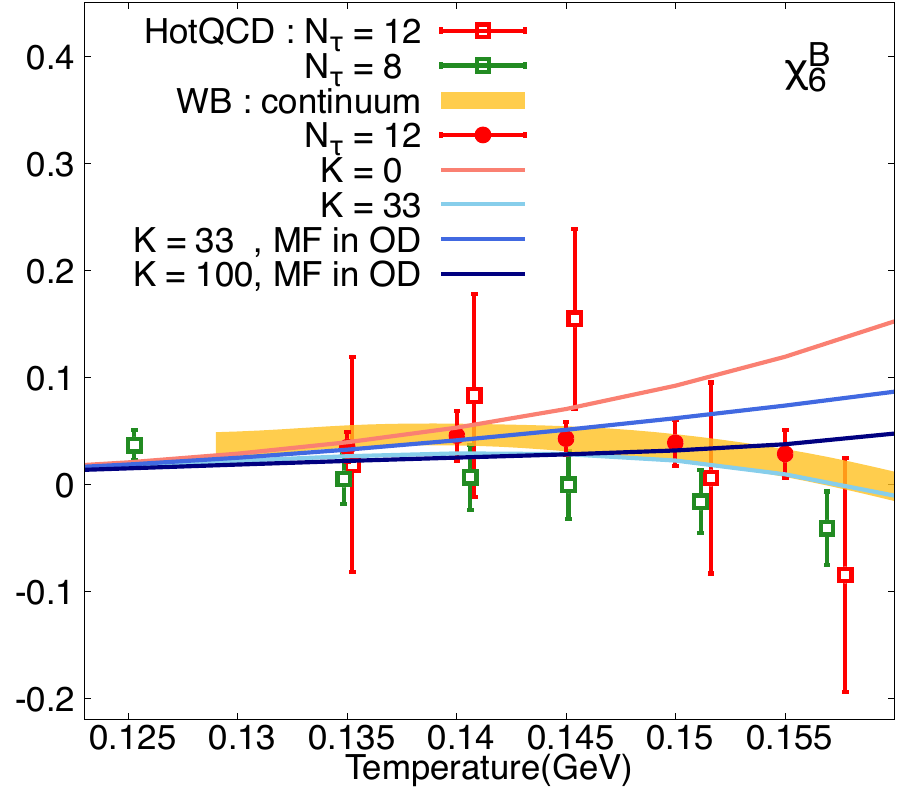}}
\subfloat{\includegraphics[width=8cm]{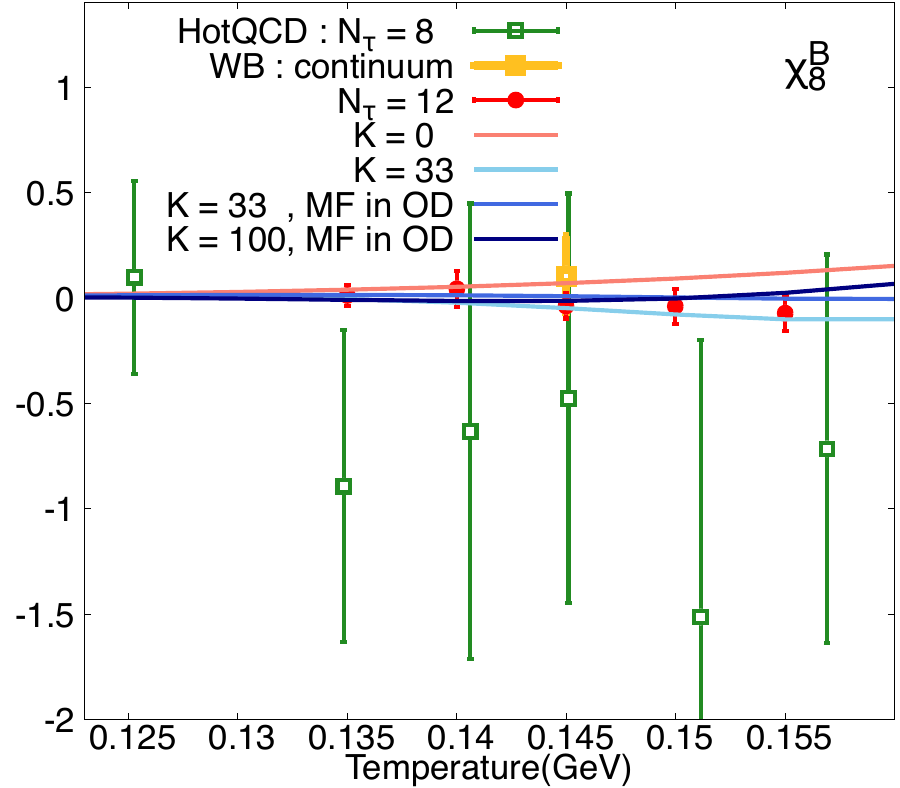}}
\caption{Top row: second-order (left) and fourth-order (right) baryon number fluctuations; bottom row: 
sixth-order (left) and eighth-order (right) baryon number fluctuations, shown as 
a function of temperature and $\mu_B=0$. The $K=0$ results represent the 
ideal QMHRG model, whereas the $K=33$ data are calculated with mean-field repulsive 
interactions among all the baryons with a mean-field coefficient of 
$K=33~{\rm GeV}^{-2}$. Similarly, MF in OD represents the case where we have included 
repulsive mean-field interaction only among the baryons in octets and decuplets. 
Lattice QCD results for $N_\tau = 8$ and $N_\tau = 12$ are shown in red and 
green colors respectively, where the squares and circles correspond 
to data from the HotQCD~\cite{Bazavov:2020bjn} and Wuppertal-Budapest 
(WB)~\cite{Borsanyi:2018grb} Collaborations. The gray and yellow 
bands represent the fluctuation data in the continuum limit from the 
HotQCD~\cite{Bazavov:2020bjn} and WB~\cite{Borsanyi:2023wno} Collaborations. }
\label{fig:Appdx_chi}
\end{figure}

Using the above expressions we calculated the fluctuations of the net baryon density up to the 
eighth order at $\mu_B=0$
and compared them to the available lattice results for different values of the parameter $K$.
The value $K=33~{\rm GeV}^{-2}$ leads to a good agreement with the lattice QCD results as one sees
from Fig. \ref{fig:Appdx_chi}. Here it is assumed that all baryons and baryon resonances contribute
to the mean field. 
If we assume that only ground-state baryons contribute to the mean field then
we do not get such a good agreement with the lattice QCD results for the same value of $K$.
However, using a larger value of $K$, namely $K=100~{\rm GeV}^{-2}$ a reasonably good agreement
with the lattice QCD results can be obtained; cf. Fig. \ref{fig:Appdx_chi}.

\section{Calculation of the equation of state within the mean-field 
HRG model}
\label{app:eos}
The energy density is defined from the pressure  as

\begin{equation}\label{epsilon}
\epsilon=-P-\beta\left(  \frac{\partial P}{\partial \beta}\right)_{\beta
\mu}.
\end{equation}

Taking the derivative of the number density with $\beta$ we find
\begin{eqnarray}
\left(\frac{\partial n_b}{\partial \beta}\right)_{\beta 
\mu}&=&\sum_{i=B} g_i \int \frac{d^3 p}{(2 \pi)^3}e^{-\beta(E_i-\mu_B+K 
n_b)}  \left[-E_i - Kn_b-\beta K  \left(\frac{\partial n_b}
{\partial \beta}\right)_{\beta \mu}	\right]  \nonumber \\
&=& -\sum_{i=B} g_i \int \frac{d^3 p}{(2 \pi)^3}f_i^b E_i - Kn_b^2-\beta
K  n_b \left(\frac{\partial n_b}{\partial \beta}\right)_{\beta \mu}.
\end{eqnarray}
Rearranging the above equation we get
\begin{eqnarray}
\left(\frac{\partial n_b}{\partial \beta}\right)_{\beta \mu}&=& \frac{ -
\sum_{i=B} g_i \int \frac{d^3 p}{(2 \pi)^3}f_i^b E_i - Kn_b^2}{1+\beta 
K  n_b}.
\end{eqnarray}
Similarly for the antibaryon, we can write
\begin{eqnarray}
\left(\frac{\partial n_{\bar{b}}}{\partial \beta}\right)_{\beta \mu}&=& 
\frac{ -\sum_{i=\bar{B}} g_i \int \frac{d^3 p}{(2 \pi)^3}f_i^{\bar{b}} 
E_i - Kn_{\bar{b}}^2}{1+\beta K  n_{\bar{b}}}.
\end{eqnarray}
The derivative of the pressure is written as
\begin{eqnarray}
-\beta\left(\frac{\partial P}{\partial \beta}\right)_{\beta \mu}&=& 
+\frac{1}{\beta}(n_b + n_{\bar{b}}) - 
\left(\frac{\partial n_b}{\partial \beta}\right)_{\beta \mu}(1+ \beta K 
n_b) -
\left(\frac{\partial n_{\bar{b}}}{\partial \beta}\right)_{\beta \mu}(1+ 
\beta K n_{\bar{b}}) \nonumber \\
&=& +\frac{1}{\beta}(n_b + n_{\bar{b}}) +\sum_{i=B} g_i \int \frac{d^3 
p}{(2 \pi)^3}f_i^b E_i + Kn_b^2 +\sum_{i=\bar{B}} g_i \int \frac{d^3 p}
{(2 \pi)^3}f_i^{\bar{b}} E_i + Kn_{\bar{b}}^2.  
\end{eqnarray}
Putting the derivative terms together in Eq.\ref{epsilon}, one can 
derive the final expression for the energy density,
\begin{eqnarray}
\epsilon &=& -\frac{1}{\beta}(n_b + n_{\bar{b}}) - \frac{K}{2}(n_{b}^2 
+ n_{\bar{b}}^2) +\frac{1}{\beta}(n_b + n_{\bar{b}}) +\sum_{i=B} g_i 
\int \frac{d^3 p}{(2 \pi)^3}f_i^b E_i +\sum_{i=\bar{B}} g_i \int 
\frac{d^3 p}{(2 \pi)^3}f_i^{\bar{b}} E_i + K(n_b^2 + n_{\bar{b}}^2) 
\nonumber \\ 
&=& \sum_{i=B} g_i \int \frac{d^3 p}{(2 \pi)^3}f_i^b E_i 
+\sum_{i=\bar{B}} g_i \int \frac{d^3 p}{(2 \pi)^3}f_i^{\bar{b}} E_i + 
\frac{K}{2}(n_b^2 + n_{\bar{b}}^2)	
\end{eqnarray}

In the ideal limit of $K=0$ the above formula reduces to the ideal HRG 
expression for energy density.
\begin{figure}[ht]
\subfloat{\includegraphics[width=6cm]{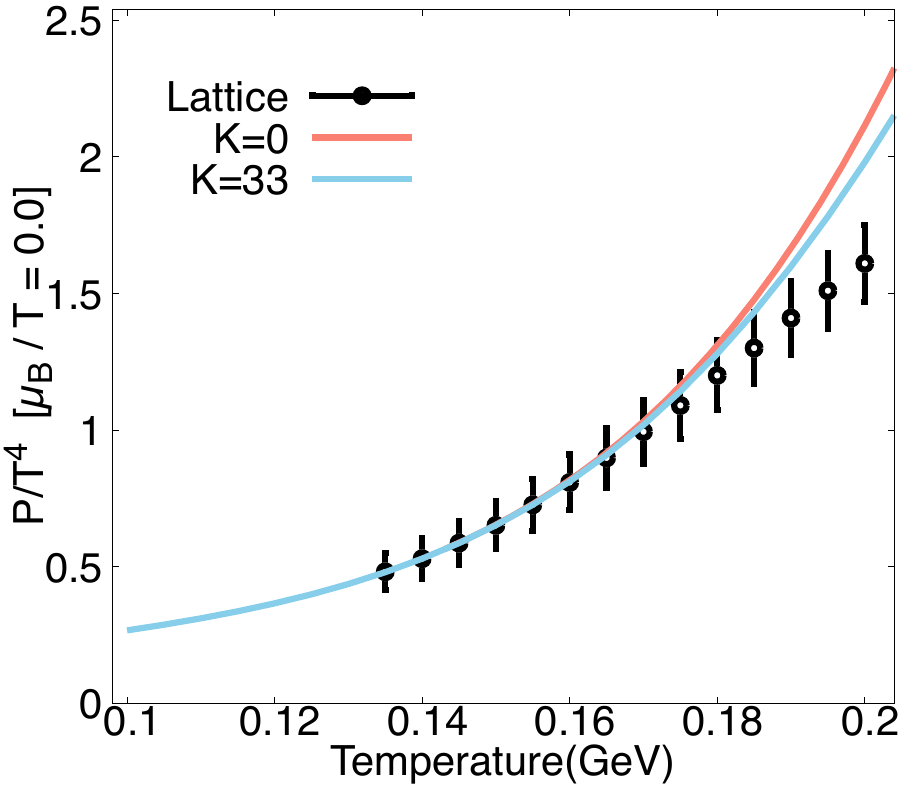}}
\subfloat{\includegraphics[width=6cm]{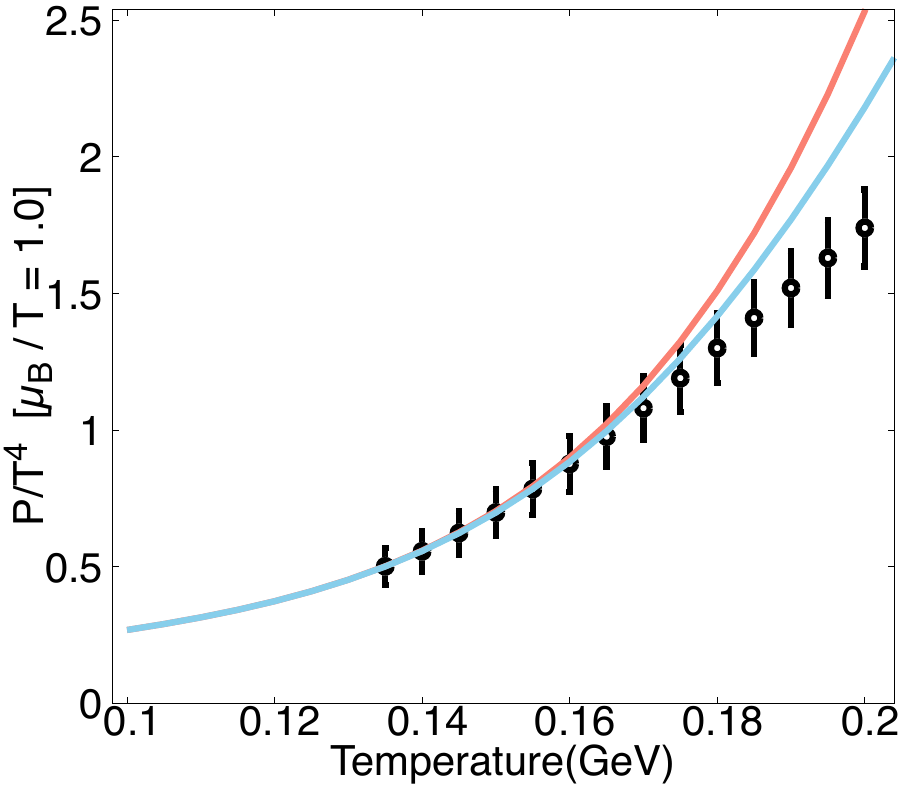}}
\subfloat{\includegraphics[width=6cm]{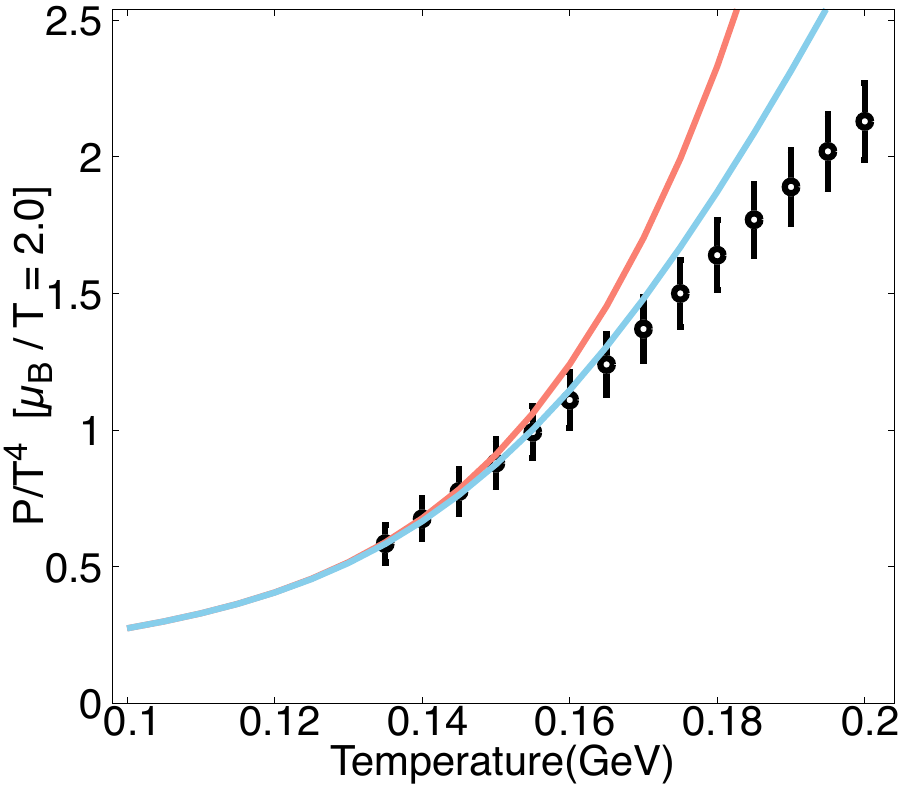}}
\\
\subfloat{\includegraphics[width=6cm]{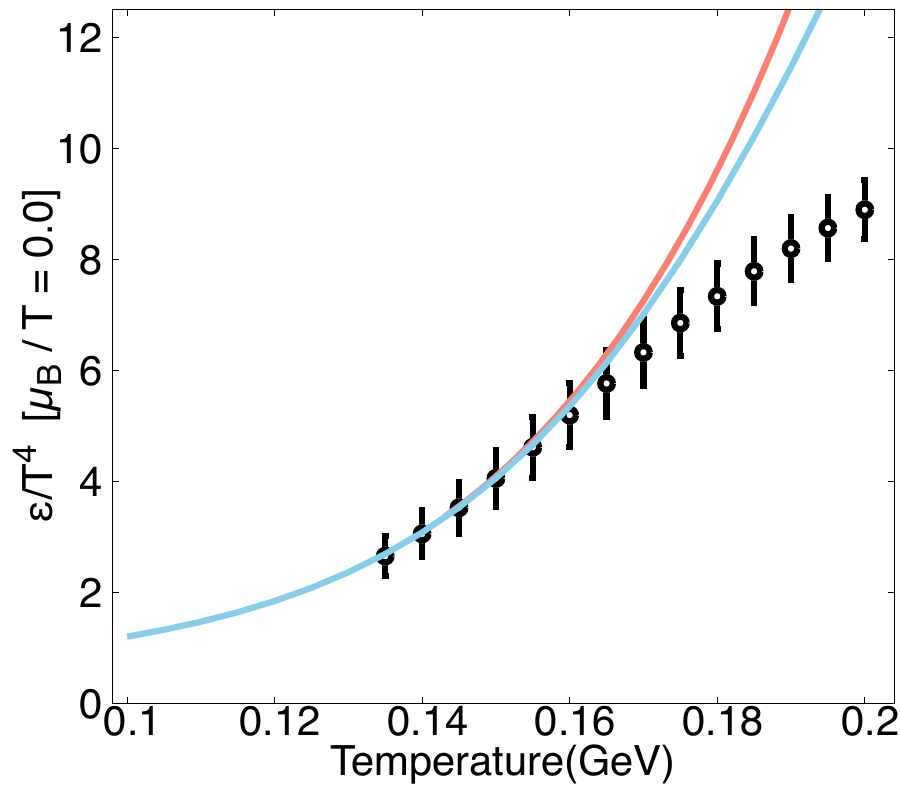}}
\subfloat{\includegraphics[width=6cm]{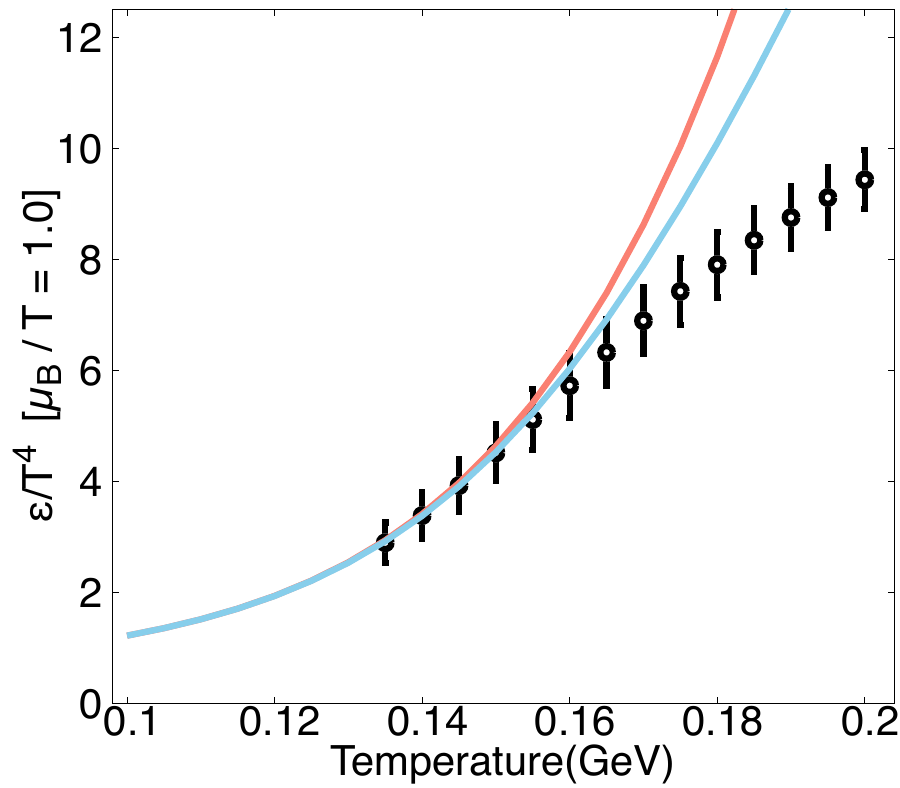}}
\subfloat{\includegraphics[width=6cm]{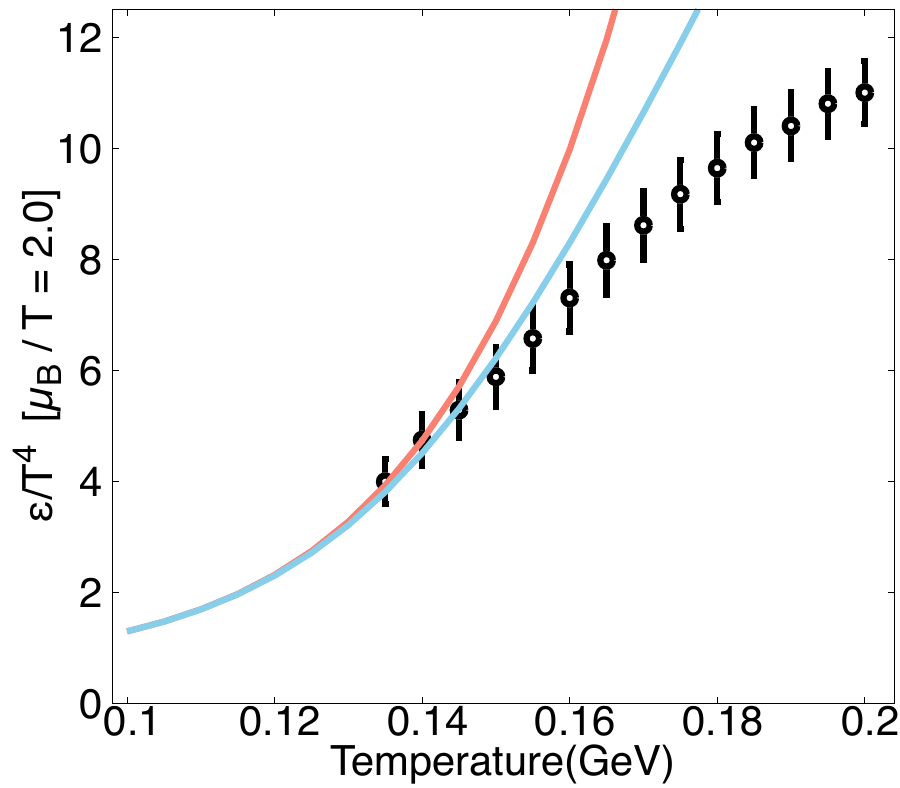}}
\caption{The pressure $P/T^4$ (upper row) and energy density $\epsilon/T^4$ 
(lower row) normalized by $T^4$ calculated for ideal (red) and 
repulsive mean-field (blue) QMHRG models. Lattice results (black points) are from 
Ref.~\cite{Bazavov:2017dus}. The results are displayed for $\mu_B/T= 0, 
1$, and $2$, in the left, middle, and right panels respectively. }
\label{fg.PT4_muB}
\end{figure}
In  Fig. \ref{fg.PT4_muB} we show the pressure and energy density in units
of $T^4$ in QMHRG and compare them with the lattice QCD results from 
Ref. \cite{Bazavov:2017dus}. The results for 
both the ideal case and repulsive mean field case with $K=33~{\rm GeV}^{-2}
$ at $\mu_B/T= 0, 1$, and $2$ are 
displayed. For all baryon chemical potentials, the QMHRG estimations 
align well with lattice results up to 150 MeV. At high $\mu_B/T$, the 
effect of repulsive mean field is more pronounced due to the increased thermal 
abundance of baryons at high density; cf. Fig. \ref{fg.PT4_muB}.

\section{Strangeness chemical potential for strange neutrality case}
\label{app:muS}

\begin{figure}[ht]
\subfloat{\includegraphics[width=8.6cm]{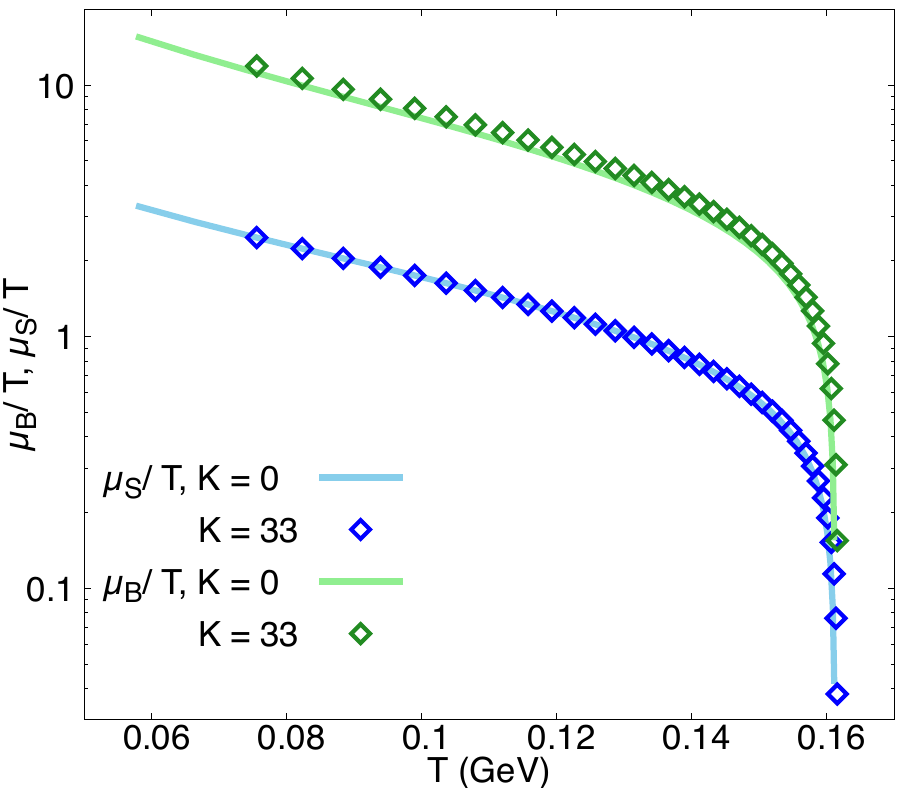}}
\subfloat{\includegraphics[width=8.6cm]{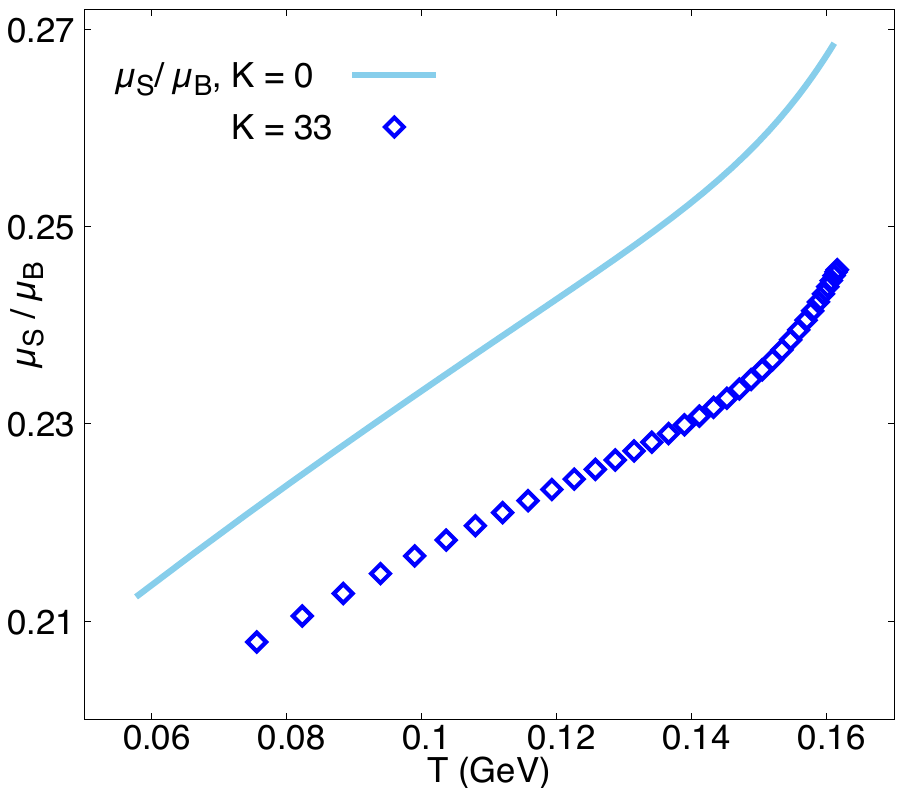}}
\caption{The chemical potentials $\mu_B/T$ and $\mu_S/T$ (left) and the ratio 
$\mu_B/\mu_S$ (right) are shown for different temperatures along the pseudocritical line. $K=0$ denotes ideal QMHRG model results, whereas $K=33$ represents the case where 
we have included repulsive interaction among all baryons with a mean-field coefficient of 
$K=33~{\rm GeV}^{-2}$. }
\label{fig:Appdx_muS}
\end{figure}

In Fig. \ref{fig:Appdx_muS} (left) we show the temperature normalized baryon 
and strangeness chemical potential along the chiral crossover line for $n_S=0$.
We see that the effect of the repulsive mean field is visible for the baryon 
chemical potential, but is very small for the strangeness
chemical potential. In Fig. \ref{fig:Appdx_muS} (right) we show the ratio $\mu_S/\mu_B$
along the crossover line. We see that the repulsive mean-field significantly reduces
this ratio compared to the ideal case.

\bibliographystyle{elsarticle-num}
\bibliography{references_paper2.bib}

\end{document}